%% file: main.tex
\documentclass[sigconf]{acmart}
\usepackage{todonotes}
\usepackage{graphicx}%
\usepackage{multirow}%
\usepackage{multicol}
\usepackage{amsmath,amsfonts}%
\usepackage{amsthm}%
\usepackage{mathrsfs}%
\usepackage[title]{appendix}%
\usepackage{xcolor}%
\usepackage{textcomp}%
\usepackage{manyfoot}%
\usepackage{booktabs}%
\usepackage{algorithm}%
\usepackage{algorithmicx}%
\usepackage{algpseudocode}%
\usepackage{listings}%
\usepackage{pifont}%
\usepackage{wrapfig}
\usepackage{enumerate}
\usepackage{acronym}
\usepackage{hyperref} 
\usepackage{relsize}
\usepackage{xspace}
\usepackage{caption}
\usepackage{subcaption}
\usepackage{pifont}
\usepackage{tabularx}
\usepackage[most]{tcolorbox}
\usepackage{hyperref}

\let\mycitep\citep
\def\cite#1{\mycitep{#1}}

\definecolor{Gray}{gray}{0.9}

\tcbset{
  gptbox/.style={
    enhanced,
    sharp corners,
    colback=gray!5,
    colframe=gray!40,
    boxrule=0.5pt,
    left=2mm,
    right=2mm,
    top=1mm,
    bottom=1mm,
    fonttitle=\normalsize\color{black},
    before skip=5pt,
    after skip=5pt,
    fontupper=\normalsize,
  },
  userbox/.style={
    gptbox,
    colback=blue!1,
    colframe=blue!15,
  }
}
\definecolor{BoxGray}{gray}{0.93}

\def\gptmodel{{GPT-5}\xspace}
\def\claudemodel{{Claude-4}\xspace}
\def\pilotmodel{{GPT-4o}\xspace}
\def\modelthree{{DeepSeek-V3.2}\xspace}
\def\modelfour{{Qwen3-Max}\xspace}

\def\prompt{\ensuremath{\texttt{P}}\xspace}
\def\promptzero{\ensuremath{\texttt{P}_1}\xspace}
\def\promptone{\ensuremath{\texttt{P}_2}\xspace}
\def\prompttwo{\ensuremath{\texttt{P}_3}\xspace}
\def\promptthree{\ensuremath{\texttt{P}_{3+}}\xspace}
\def\promptpilot{\ensuremath{\texttt{P}_{3pilot}}\xspace}

\def\code{\ensuremath{c}\xspace}
\def\annotator{\ensuremath{a}\xspace}
\def\Totalprompts{\ensuremath{\#P}\xspace}
\def\Totalcodes{\ensuremath{\#C}\xspace}
\def\Totalannotators{\ensuremath{\#A}\xspace}
\def\TotalLLMs{\ensuremath{\#L}\xspace}
\def\Totalcomments{\ensuremath{N}\xspace}

\def\StarHallucinate{\ding{55}}
\def\StarUnusable{\ding{72}\xspace}
\def\StarLimited{\ding{72}\StarUnusable}
\def\StarModerate{\ding{72}\StarLimited}
\def\StarSubstantial{\ding{72}\StarModerate}
\def\StarAlmostPerfect{\ding{72}\StarSubstantial}

\newcommand{\ea}{{et al.}\xspace}

\copyrightyear{2026}
\acmYear{2026}
\setcopyright{cc}
\setcctype{by}
\acmConference[ICPC '26]{34th IEEE/ACM International Conference on Program Comprehension}{April 12--13, 2026}{Rio de Janeiro, Brazil}
\acmBooktitle{34th IEEE/ACM International Conference on Program Comprehension (ICPC '26), April 12--13, 2026, Rio de Janeiro, Brazil}
\acmDOI{10.1145/3794763.3798172}
\acmISBN{979-8-4007-2482-4/2026/04}

\begin{document}

\onecolumn
\input{cover}

\twocolumn
\input{body}

\bibliographystyle{ACM-Reference-Format}
\bibliography{biblio}

\end{document}

%% file: cover.tex
\newenvironment{project}[1]
{\par
 \bigskip
 \begin{wrapfigure}{l}[0pt]{1in}
 %\vspace{-15pt}
 \includegraphics[width=1in,clip,keepaspectratio]{#1}
 \vspace{-25pt}
 \end{wrapfigure}
 \footnotesize \noindent}
{\par\bigskip}

\newenvironment{bio}[1]
{\par
 \bigskip
 \begin{wrapfigure}{l}[0pt]{0.5in}
 \vspace{-15pt}
 \includegraphics[width=0.5in,keepaspectratio]{#1}
 \end{wrapfigure}
 \footnotesize \noindent}
{\par\bigskip}

\begin{figure}
    \centering
    \includegraphics[width=.3\textwidth]{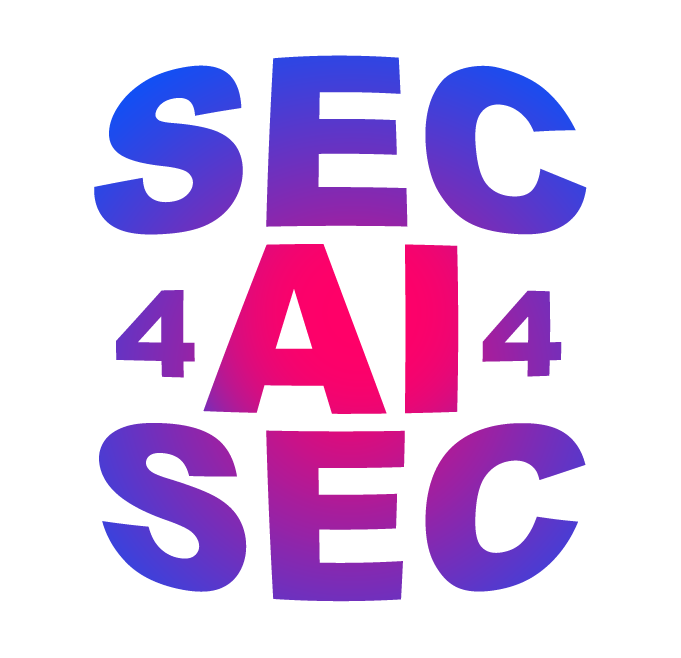}
\end{figure}

\vspace{2\baselineskip}

%\begin{figure}[h!]
%    \centering
%    \includegraphics[width=\textwidth]{logos/sec4ai4sec.png}
%\end{figure}

\vspace{2\baselineskip}

\begin{center}
{\huge \textbf{LLMs for Qualitative Data Analysis Fail on Security-specific
Comments in Human Experiments}}
\end{center}

\vspace{\baselineskip}

{\large
Authors:
\begin{itemize}
    \item[]\textbf{Maria Camporese}, University of Trento (Italy)
    \item[]\textbf{Fabio Massacci}, University of Trento (Italy), Vrije Universiteit Amsterdam (The Netherlands)
    \item[]\textbf{Yuanjun Gong}, University of Trento (Italy)
\end{itemize}
}

\vfill

\begin{figure}[h]
\vspace{-\baselineskip}
\includegraphics[height=3cm]{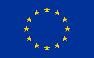}
~\includegraphics[height=3cm]{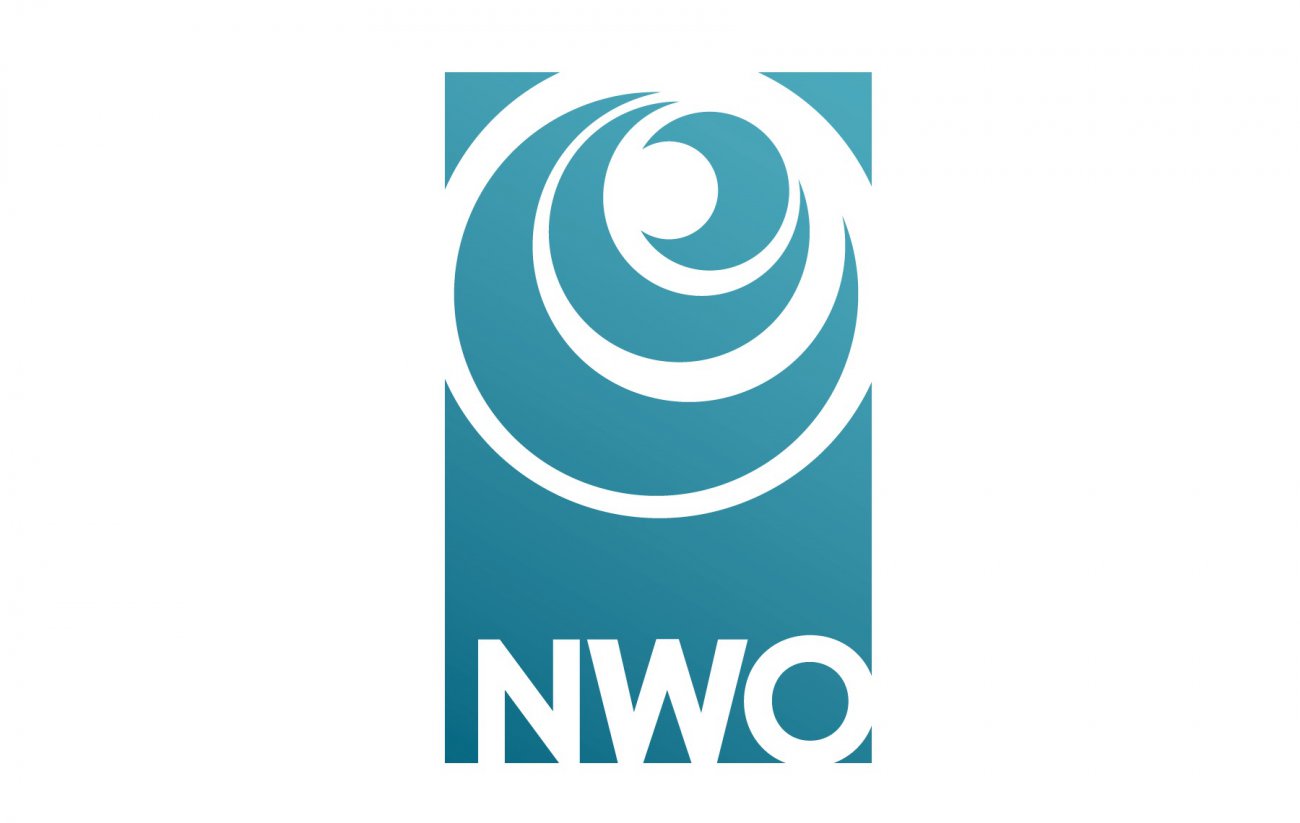}
~\includegraphics[height=3cm]{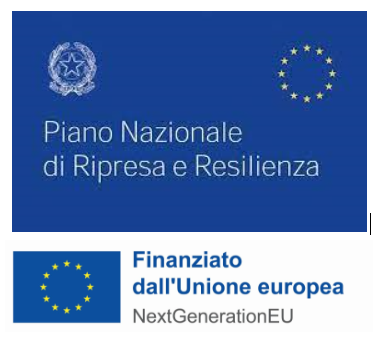}

\end{figure}

\noindent This work has been partly supported by the European
Union (EU) under Horizon Europe grant n
. 101120393
(Sec4AI4Sec), by the Nederlandse Organisatie
voor Wetenschappelijk Onderzoek (NWO) under grant n.
KIC1.VE01.20.004 (HEWSTI), and by the Italian Ministry of University and Research (MUR), under the P.N.R.R. – NextGenerationEU grant
n. PE00000014 (SERICS). This paper reflects only the author's view and the funders are not responsible for any use that may be made of the information contained therein.

\clearpage

% \vspace*{-6\baselineskip}
\begin{project}{logos/EU-Logo.png}
\textbf{Cybersecurity for AI-Augmented Systems (Sec4AI4Sec)} . As artificial intelligence (AI) becomes omnipresent, even integrated within secure software development, the safety of digital infrastructures requires new technologies and new methodologies, as highlighted in the EU Strategic Plan 2021-2024. To achieve this goal, the EU-funded Sec4AI4Sec project will develop advanced security-by-design testing and assurance techniques tailored for AI-augmented systems. These systems can democratise security expertise, enabling intelligent, automated secure coding and testing while simultaneously lowering development costs and improving software quality. However, they also introduce unique security challenges, particularly concerning fairness and explainability. Sec4AI4Sec is at the forefront of the move to tackle these challenges with a comprehensive approach, embodying the vision of better security for AI and better AI for security. More information at \textbf{\url{https://sec4ai4sec.eu}}.
\end{project}

\begin{project}{logos/NWO-logo.jpg}
\textbf{Hybrid Explainable Workflows for Security and Threat Intelligence (HEWSTI)} In research into threats to safety and security, people and AI collaborate to obtain actionable intelligence. Their sources and methods often have significant uncertainties and biases. Experts are aware of these limitations, but lack the formal means to handle these uncertainties in their daily work. This project will invent a ‘metadata of uncertainty’ for threat intelligence (in both machine-readable and also human-interpretable forms) and validate it empirically. Intelligence agencies will then be able to explicitly consider the trade-off between the accuracy, proportionality, privacy, and cost-effectiveness of investigations. This will contribute towards the responsible use of AI to create a safer, more secure society.
\end{project}

\begin{project}{logos/pnrr-next_generation-logo.png}
\textbf{In searCh Of eVidence of stEalth cybeR Threats
(COVERT)} AT 3 aims to analyze emerging attack methodologies and develop advanced methods for detecting attacks and identifying guidelines for designing IT systems that ensure reduced vulnerability to new attack categories. The detailed objectives can be divided into four macro categories: (i) Development of advanced tools for analyzing malware and software aimed at identifying vulnerabilities that could be exploited by malware; (ii) Development of tools for analyzing network traffic to identify communications related to ongoing attacks; (iii) Development of machine learning systems that are robust to attacks and through which it is possible to extract knowledge aimed at creating more advanced tools for timely analysis and early identification of attacks; (iv) Analysis of the "human factors" involved in an attack with the development of tools for analyzing and correlating information from OSINT (open sources intelligence) and for the defense and prevention of attacks based on social engineering techniques.
\end{project}

\vfill

\begin{bio}{photos/camporese}
\textbf{Maria Camporese} (MSc 2022) is a PhD student at the University of Trento, Italy. Her research interests include security and machine learning. Contact her at \emph{maria.camporese@unitn.it}.
\vspace{0.8cm}
\end{bio}

\begin{bio}{photos/massacci}
\textbf{Fabio Massacci} (Phd 1997) is a professor at the University of 
Trento, Italy, and Vrije Universiteit Amsterdam, Fabio Massacci is a professor at the University of Trento, Trento, Italy, and Vrije Universiteit Amsterdam, 1081 HV Amsterdam, The Netherlands. His research interests include empirical methods for the cybersecurity of sociotechnical systems. For his work on security and trust in sociotechnical systems, he received the Ten Year Most Influential Paper Award at the 2015 IEEE International Requirements Engineering Conference. He is named co-author of CVSS v4. He leads the Horizon Europe Sec4AI4Sec project and the Dutch National Project HEWSTI.  He is past chair of the Security and Defense Group of the Society for Risk Analysis, and IEEE CertifAIEd Lead Assessor. Contact him at \emph{fabio.massacci@ieee.org}.
\vspace{-0.2cm}
\end{bio}

\begin{bio}{photos/gong}
\textbf{Yuanjun Gong} (PhD 2025) is a postdoc researcher at the University of Trento, Italy. Her research interests include static analysis,
software security and machine learning. Contact her at \emph{yuanjun.gong@unitn.it}.
\vspace{1cm}
\end{bio}

How to cite this paper:
\begin{itemize}
    \item Camporese, M., Massacci, F. and Gong, Y. LLMs for Qualitative Data Analysis Fail on Security-specific Comments in
Human Experiments. \emph{Proceedings of the 2026 IEEE/ACM 34st International Conference on Program Comprehension (ICPC)}. IEEE Computer Society.
\end{itemize}

License:
\begin{itemize}
\item This article is made available with a perpetual, non-exclusive, non-commercial license to distribute.
%\item The graphical abstract is an artwork by Anna Formilan.
\end{itemize}

\clearpage

%% file: body.tex
\title{LLMs for Qualitative Data Analysis Fail on Security-specific Comments in Human Experiments}

\author{Maria Camporese}
\affiliation{%
  \institution{University of Trento, IT}
  %\city{Trento}
  \country{}
}
\email{maria.camporese@unitn.it}

\author{Fabio Massacci}
\affiliation{%
  \institution{University of Trento, IT,\\ Vrije Universiteit Amsterdam, NL}
  % \city{Trento}
  % \country{Trento, Italy and Amsterdam, Netherlands}
  \country{}
}
\email{fabio.massacci@ieee.org}

\author{Yuanjun Gong}
\affiliation{%
  \institution{University of Trento, IT}
  %\city{Trento}
  \country{}
}
\email{yuanjun.gong@unitn.it}

\renewcommand{\shortauthors}{Maria Camporese et al.}


\begin{abstract}
% \todo[inline]{Fabio: Revised, looks good. I addede four instead of two}
[\emph{Background}:] Thematic analysis of free-text justifications in human experiments provides significant qualitative insights. 
Yet, it is costly because reliable annotations require multiple domain experts. Large language models (LLMs) seem ideal candidates to replace human annotators.
[\emph{Problem}:] Coding security-specific aspects (\textit{code identifiers mentioned}, \textit{lines-of-code-mentioned}, \textit{security keywords mentioned}) may require deeper contextual understanding than sentiment classification.
[\emph{Objective}:] explore whether LLMs can act as automated annotators for technical security comments by human subjects.
[\emph{Method}:] We prompt four best LLMs on LiveBench to detect nine security-relevant codes in free-text comments by human subjects analyzing vulnerable code snippets. Outputs are compared to the human annotators along with Cohen's Kappa (chance-corrected accuracy). We test different prompts mimicking annotation best practices: emerging codes, a detailed codebook with examples, and conflicting examples.
[\emph{Negative Results}:] We observed marked improvements only with the code descriptions, but they are not uniform across codes and not sufficient to reliably replace a human annotator.
[\emph{Limitations}:] Additional studies with more LLMs and annotation tasks are needed.
\end{abstract}

\keywords{Qualitative Data Analysis, Thematic Analysis, Large Language Models, Security-Specific Annotation}

\maketitle

\section{Introduction}\label{sec:intro}
% \todo[inline]{Fabio: Revised, looks good.}
Who has run a software engineering experiment with human subjects and not received a reviewer's comment ``You should manually analyze the participants' explanations''? 

The process for doing this qualitative analysis is called Thematic Analysis (TA) \cite{dai2023llm}. 
TA is a method for identifying, coding, and interpreting patterns of meaning (themes) within qualitative data \cite{clarke2017thematic}. The aim is to distill insight from large volumes of qualitative material to allow for quantitative analysis \cite{zhang2009qualitative}. 
Fragments of textual, audio, or video recordings are annotated with labels called \emph{codes}. Codes represent the smallest meaningful units, capturing noteworthy aspects of the recorded material. These codes are used for constructing broader themes, i.e.\ recurring patterns of meaning.
To ensure reliable analysis, the same piece of data is assigned to at least one coder and one reviewer. Further, human coders develop and deepen their data interpretation and coding over multiple iterations, making TA labor-intensive and time-consuming. 

In the realm of empirical software and security engineering, thematic analysis is used in a variety of settings. 
For example, opinion mining over software repositories' comments is a popular research area \cite{lin2022opinion}. 
In experiments with humans, qualitative analysis is used to validate the rationale behind their choices \cite{papotti2024acceptance,papotti2025effects}.

When discussing coding for opinion mining, one typically focuses on the Sentiment polarity identification (e.g., positive, neutral, or negative \cite{lin2022opinion}), viewpoints and perspectives identification (e.g., general reasons behind forking a repository \cite{jiang2017and}), or other knowledge extraction (e.g.\ usefulness of code reviews \cite{rahman2017predicting}). This can be done both manually or automatically, e.g., the cited work by Jiang \cite{jiang2017and} had both a manual annotation of open-ended questions and automated analysis of developer responses.  More recent works proposed to use LLMs \cite{dai2023llm,maehlum2024s}.
Using LLMs for replacing at least one human annotator seems, therefore, a low-hanging fruit. 

We are interested in more specific, \emph{security-relevant annotations} in which the LLM has to extract more domain-specific answers, such as whether the vulnerability type is mentioned in the human explanations, or whether a possible exploit is mentioned, or some security keywords are mentioned. They are further detailed in Table \ref{tab:input}. Detection of such qualitative codes has also used by SAP to build their heuristic approach to identify security fixes among thousands of commits starting from vulnerability descriptions \cite{sabetta2024known}.

We therefore raise two research questions to explore the potential of LLMs for automatic technical comment thematic annotation:
\begin{description}
    \item[RQ1.] Can LLMs replace a human annotator in \emph{security technical annotations}?
    \item[RQ2.] What is the impact of different prompts mimicking the best practice of code refinements on LLMs' performance on \emph{security technical annotations}?
\end{description}

% \subsection{The Contribution of This Paper}

We propose and instantiate an experimental methodology where LLMs receive a dataset of human comments, partially annotated by other humans, and must generate the annotations of security-relevant codes for a fraction \emph{not yet} annotated comments (the ones assigned to the replaced human). The design of the prompting strategy reflects the human coding procedure (emerging coding at first, followed by a detailed codebook with examples, finally addressing conflicting examples). 

For the actual experiments, we then identified four state-of-the-art LLMs and asked them to generate the security-relevant codes instead of human annotators. When looking for a dataset for annotations, we realized that LLMs are likely trained on public datasets. Hence, we reached out to the authors of a security study where security technical comments have been manually annotated \cite{papotti2024acceptance,papotti2025effects} and asked if they had additional, not yet published annotations. They provided a thematic analysis of a different, unpublished experiment in which they had multiple annotators. This provided us with a dataset of $n=263$ human-generated comments where 13 security-relevant codes ($c=9$ security-relevant codes are tested) have been annotated by four people (see some examples in Table\ref{tab:input}).

Two key observations are important to evaluate such replacement: first, we should not hold an LLM to a higher standard than a human annotator. Second, for a code to be useful, it has to be at once recurrent but sparse to truly capture different concepts \cite{armborst2017thematic}. Hence, an LLM could obtain a good accuracy just by chance because a meaningful code is typically not applicable to a random comment. So we measure whether the replaced LLM can achieve the same Cohen's Kappa as the (replaced) human annotator and their reviewer.  The same formula used for Cohen's Kappa can also be interpreted as chance-corrected accuracy \cite{barnston1992correspondence}, a metric proposed in weather forecasting to test for rare events. Indeed, when using traditional accuracy, we obtained over 75\% in our experiment, when correcting for chance, the results dropped to half that value.

\begin{tcolorbox}[colback=gray!3, colframe=black!30, boxrule=0.4pt, arc=1mm]
\textbf{Negative Result:} LLMs cannot reliably replace human annotators for security-relevant annotations, not even with a detailed code-book and examples from other annotators.
\end{tcolorbox}

% \subsection{Plan of the Paper}
%The remainder of this paper is structured as follows.

% \todo[inline]{YG: change it according to the current structure}
% Section \ref{sec:methodology} details our methodology, including the research questions, problem specification, study design, data collection process, input preparation, annotation generation, prompt refinement, and analysis approach.
% Section \ref{sec:data:collection} describes the datasets used for experimentation and the preparation steps undertaken to ensure reliability and comparability.
% Section \ref{sec:results} presents the experimental results, answering RQ1 on the ability of LLMs to replace a human annotator, and RQ2 on the impact of prompt engineering.
% Section \ref{sec:related:work} reviews related work in three areas relevant to our study: opinion mining for software development, qualitative data analysis in software engineering (SE) experiments, and the emerging use of large language models (LLMs) for qualitative research.
% Section \ref{sec:discussion:threats} discusses the implications of our findings for the use of LLMs in domain-specific qualitative annotation and the threats to validity, including the measures taken to mitigate them.
% Finally, Section \ref{sec:conclusions} concludes the paper and suggests directions for future work.

\subsection{Artifact Availability}
The data and the code implementation are open-sourced in \url{https://zenodo.org/records/18742065}.

% \subsection{Artifact Availability Statement}

% \input{EMSE/Submission/data-statement}

% \section{Background}

\section{Human Annotations - Problem Specification}\label{sec:problem}
% \paragraph{Problem Specification}
% \fm{List of comments generated by each participants and a list of codes that could be applied to each comment as decided by a human annotator. Each code also has a comment. Expand the problem specification with experiment 1.}

Our research focuses on one of the simplest protocols for generating reliable annotations, as shown in Figure~\ref{fig:problem}.
In the qualitative data collection from Papotti \ea \cite{papotti2025effects}, the authors describe the process they used for annotation, which is quite standard for applied Thematic Analysis~\cite{guest2011applied, gregory2015paradoxes}: first, a subset of authors jointly reviewed a sample of justifications for each vulnerable scenario to identify a first set of emerging codes; second, the codes are consolidated into a codebook (Table 8 in \cite{papotti2025effects}). After an additional phase of consultation, the author marked all remaining justifications, and an independent researcher checked them. Finally, additional conflicts were discussed by a subset of the authors and resolved. 

With human annotators $(A, B, C, D ,...)$, the procedure should be split down into four steps:
\begin{figure}[t]
    \centering
\includegraphics[width=0.95\linewidth]{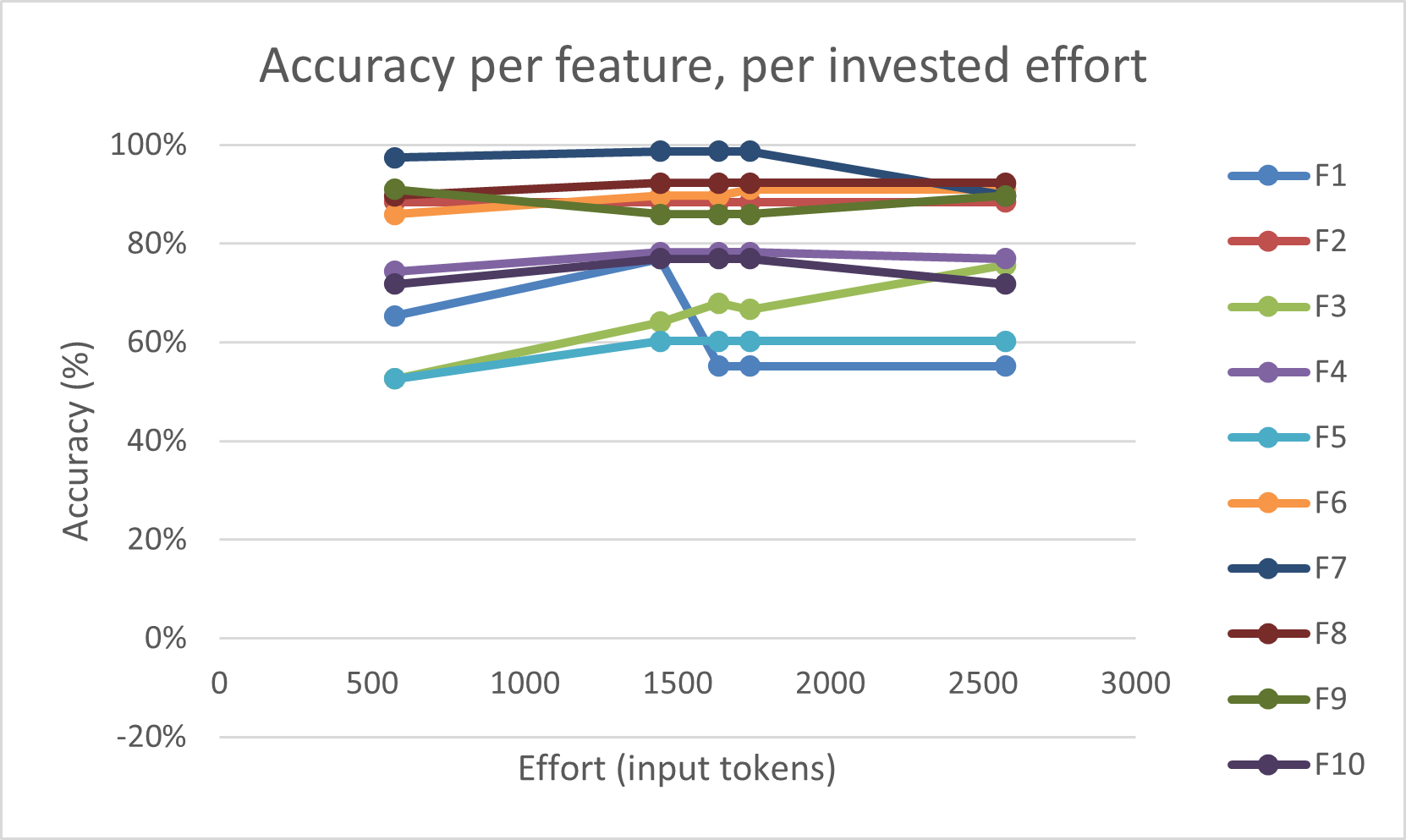}
% \todo[inline]{YG: Adjust/revise figure 1 (prompts)}
    \caption{Annotation by human annotators and LLMs}
    \label{fig:problem}
    %The text below is invisible is used 
    %by ACM to generate the description 
    %for blind people
    \Description{This figure shows the four stages of the human annotation process.}
\end{figure}

\begin{table*}[t]
\caption{Example of Codes and Related Comments}\label{tab:input}
    \begin{tabular}{cp{0.17\linewidth}p{0.41\linewidth}p{0.28\linewidth}}
    \hline    
    \textbf{Abbr.} &
    \textbf{Code Definition} &
    \textbf{Code applicable Comments} &
    \textbf{Code inapplicable Comments} \\ \hline
    Var & Variable/Method identifiers are mentioned  & I think it had something to do with inv and debug. & header buffer capacity is not checked.
\\ \hline 
    Lin & Lines number mentioned  & In this line, the memory at constant address is copied, this memory will most likely not stay consistent among systems. & The null value may cause the function crash.\\ \hline 
    Key & Relevant Keywords mentioned  & If the file attribute in the code is tampered with for an instance of the BlockDriverState it might happen that the code runs an infinite recursion, essentially stopping the machine. & I don't see any vulnerabilities.
\\ \hline 
    Vul & Text is related to the specific vulnerability  & Might modify environment variable with strtol. & I think that negative checkin triggers the vulnerability for the ML, not sure. \\ \hline 
    Sec & Text related to security  & A check should be added which checks if the buffer is not overflowing. & the datatype of debug wont support strtol.\\ \hline
    Exp & Potential exploit is mentioned  & The memcpy() function might be injected with a size that is too big for the program to handle, therefore freezing the computer's resources. & stderr is not created.
\\ \hline
    %\#B3 & The ML choice/ interpretation is mentioned  & The other line had only highlighted a semi colon which is misleading \\ \hline
    %\#C4 & ML is confused  & The ML algorithm might not "understand" what the g\_assert function does, and it will assume that it's vulnerable just because you have a pointer reference/memory operation \\ \hline
    %\#B6 & Highlights are misleading  & Though I think the code is not vulnerable but the highlighted lines might be interpreted as vulnerable by a machine learning tool. \\ \hline
    %\#C6 & No highlights, hard to guess which part ML found vuln  & Without highlights is hard to guess what the ML found faulty, especially since I did not think that it was vulnerable. \\ \hline
    NoVul & I don't find vulnerability  & Since I did not find the code to be vulnerable, I have not marked any sections. The confidence however is low, as I am struggling to find any vulnerable line. & I am not sure. \\ \hline
    Unsr & I'm not sure  & The written buffer may overflow like said before, but I'm not really sure.& memcpy buffer overflow.
\\ \hline
    Unclr & Unclear answer  & Address. & peer might be null.\\ \hline
    \end{tabular}
\end{table*}

\begin{enumerate}
\setcounter{enumi}{-1} 
\item Emergence of codes
\begin{enumerate}
    \item  \label{step:emergence}  Annotators individually read some samples and make up their own codes (principle of emergence). 
    \item \label{step:define:names} The annotators gather up the codes and agree on code names (Column 1 of Table\ref{tab:input}).
\end{enumerate}
\item First annotation with shared codes (names)
\begin{enumerate}
    \item \label{step:annote:names} The annotators get an initial understanding from the background information and the codebook. The annotators individually mark a large fraction of their comments\footnote{The golden standard is that all comments should be annotated to avoid influencing each other. In the experience of one author, annotators tend to stop and call a meeting when too many cases accumulate in which the annotator is unsure.}. 
    \item \label{step:define:definition}At this point, a formal definition of the codebook is agreed (Column 1 and 2 of Table\ref{tab:input}). 
\end{enumerate}
\item Completed annotation with codebook and first review
\begin{enumerate}
   \item\label{step:annote:definition}The annotators get a deeper understanding of the task from the explicit definition for every code;  They return to their comments and consolidate their codes on all assigned comments. 
    \item \label{step:review} Each annotator reviews the annotations from another annotator, marking agreement and conflicts. Reviewers are randomly assigned.\footnote{To reduce cost, one might consider a single annotator and a single reviewer, but this choice reduces the reliability too much. Also in this scenario, the human reviewer would have to review all comments anyhow, and then they could directly mark them.} 
    \item \label{step:define:dodont} By the end of this step, the disagreements are discussed and Column 1-4 of Table\ref{tab:input} are generated as examples of representative do and don't. 
\end{enumerate}
\item Final consolidate coding with do-s and don't-s
\begin{enumerate}
    \item \label{step:annote:dodont}The annotators get a comprehensive understanding of the task from the examples of other annotators, as well as the conflicting samples; Annotators individually finalize the annotations along the codebook, bearing in mind the conflicting examples. 
        \item 
By the end of this step, the final review discuss and collectively resolve all standing disagreements.
\end{enumerate}
\end{enumerate}
For the application scenario, our research focuses on the \emph{technical comments} annotation. The technical comments are provided by human subjects in response to a request to perform a security analysis of code snippets or patches. The \emph{background information} is the summary information on the type of vulnerabilities and evidence shown to the commenter and known to the annotator. The \emph{codes} are thematic aspects applied to these comments. The annotators are expected to label a code as ``1'' if the code applies to the participant's comment and ``0'' if not.

Table\ref{tab:input} describes the definition of codes and some representative examples illustrating the annotation tasks, including positive samples and negative samples. 
For example, in the comment ``I think it had something to do with inv and debug'', the code ``Variable/Method identifiers are mentioned'' (abbreviated as \textit{Var}) is labeled as ``1'', since the comment mentions the identifier ``inv''. 

In our dataset, the four annotators divided the comments into four sets and followed the protocol above, each of them randomly annotating a fourth of the comments and reviewing the annotations of a different annotator.

The LLM should start after step~\ref{step:define:names}, when at least the code names are defined by brainstorming among humans. It should then be able to replace any of the human annotators (not as a reviewer), with an acceptable success in each of the annotation steps (\ref{step:annote:names}, \ref{step:annote:definition}, \ref{step:annote:dodont}) using the corresponding information resulting from the previous steps (\ref{step:define:names}, \ref{step:define:definition}, \ref{step:define:dodont}).
To apply the LLM support, we follow the process of the cognition progress of human annotators, and design the prompts to provide the corresponding information to the LLM. \emph{If} the LLM has the ability to substitute human annotators, it should have a similar performance after getting the same prompt they got.

For the last steps (\ref{step:annote:definition}) we consider also a \emph{favourable version of the task} for the LLM, as we show in the first half of the bottom part of Figure \ref{fig:problem}: we give it the comments of the other annotators who later would be reviewers of its own task. Finally, for step \ref{step:annote:dodont}, we ask it to annotate only the comments on which the humans agreed, while all comments on which humans disagreed are provided as examples. Both `helping hands' should actually boost both Cohen's Kappa and the chance-corrected accuracy.

This favorable condition can be acceptable if the LLM is intended to replace the N-th annotator after the original N-1 humans have already run the protocol on their part and want to delegate the completion of the remaining 1/Nth part of the comments to the LLM. In this scenario, it makes no sense to withhold from the LLM the consensus they have already reached. Still, the LLM is favored as in the actual replacement scenario, it would also have to mark the difficult comments on which there was a human disagreement.

% \todo[inline]{FM: I think this is enough for being a \emph{problem description}. All the text below should be moved to the methodology.}

\section{Methodology}\label{sec:methodology}

The next section provides a broad overview of the main stages of our study to address our research questions. Some additional steps are further refined, including data collection, input preparation, annotation generation, prompt refinement, and result analysis.

\subsection{Methodology Overview}

\paragraph{Data Collection and human annotation.} 

We initially planned to use data from published experiments (e.g. \cite{papotti2025effects}) when we realized that annotations published on the internet might be known to the LLM. Other experiments on commits and code analysis either did not make available the codes or were not on security relevant (e.g. \cite{jiang2017and,papotti2024acceptance}). 
So we contacted the authors of \cite{papotti2024acceptance,papotti2025effects} and obtained a different dataset that is not yet published. 
The human annotation procedure is done by four annotators, following the process in \S~\ref{sec:problem}. 
The dataset is represented as a spreadsheet and contains the labels of code snippets or patches, the corresponding comments, and the annotator-labeled codes. 
We also obtain the reviewer-corrected code for each annotated comment (which we used as the final ground truth in the best prompt). The security-specific comments collection and the annotation will be further described in \S\ref{sec:sec:data:collection}.  

\paragraph{Input Preparation.}
To be consistent with the human protocol, the input to LLMs is the same spreadsheet where the codes labeled by the selected human annotator are cleared. The remaining information in the spreadsheet, such as the codes of the remaining annotators, can provide extra information for the LLMs to understand the task. As mentioned in Section~\ref{sec:problem}, this is a significant help for the LLM: the human annotators had to perform the work independently. We believe it is justified as the idea is to use the LLM to ``finish off'' the work for the first $\Totalannotators-1$ annotators

\paragraph{Prompt Refinement.} 
To guide the model effectively, imitating the human annotation process (\S\ref{sec:problem}), details and examples are progressively added to the prompts (research goal and experimental design, and format and semantics of the spreadsheet used for annotation, examples of both successes and failures).
A more detailed explanation is provided in \S\ref{sec:sec:prompting}.

\paragraph{LLM Annotation Generation.}
The LLM under evaluation will be instructed by the prompts to fill in the empty cells of the input spreadsheet generated from the previous step. The LLM will be prompted to identify the present or absent labels of each code in each participant's comment, and generate the annotations based on the remaining information from the spreadsheet and the knowledge provided by the prompts. The result of LLM auto-annotation should also be reported in a spreadsheet, generated by LLMs.

\paragraph{Analysis.} 
The LLM annotation experiments will be conducted individually, each time employing one \TotalLLMs LLMs and one of the \Totalprompts prompt types to replace one of the \Totalannotators annotators. The results for each run will be presented across \Totalcodes codes. The annotations from LLMs are analyzed against the annotators' and the reviewers' annotations, to compute the suitable performance metrics (most notably Cohen's $\kappa$) and identify statistically significant differences among LLMs or prompts. The analysis methods are refined in \S\ref{sec:sec:analysis}.

\subsection{Comments Collection and Annotation}
\label{sec:sec:data:collection}
% \todo[inline]{YG: to avoid dataset2 as an expression, and to address that it is small but private}
% \yg{We need to report how many comments we have in each dataset, how many participants, annotators, how many conflict samples are used in the prompt and how many samples are to annotated by LLMs.}

% To avoid any confirma-895
% tion bias the coders only looked at the information on the type of vulnerability896
% (e.g. XSS vs Path Traversal) without knowing which lines were formally in-897
% serted in the response, and whether the lines proposed by the participant were898
% actually vulnerable or not according to the ground truth.

A realistic dataset requires underlying software engineering experiments with raw comments of the human participants \emph{and} the code labels from human annotators. 
The natural production of the data is that, in an underlying software or security engineering experiment, the participants were asked to give security-related comments on the provided code snippet or patches to explain the rationale for the choices of the experiment. Then, the human-annotators analyzed the comments following the steps in \S\ref{sec:problem}.

\paragraph{The Underlying Experiment.} The annotation dataset used in this research came from an unpublished experiment from the researchers behind Papotti \ea \cite{papotti2025effects}. In their experiment, a group of researchers conducted a controlled experiment to assess whether AI-generated suggestions could support Computer Science Master's students in identifying vulnerabilities within source code snippets. In that experiment, participants were shown six code snippets, each in one of three possible versions: a plain (control) version, or a version enhanced with highlights generated via explainability methods. 
The highlights correspond to codes deemed most relevant by VulDeePecker~\cite{li2018vuldeepecker}, a DL-based vulnerability detection model. The highlighted versions were generated using either a white-box method (Gradient $\times$ Input~\cite{zeiler2014visualizing}) or a black-box method (LEMNA~\cite{Guo2018LEMNAED}). 
The code snippets included common vulnerability types: buffer overflows, format strings, and NULL pointer dereferences.

The experimenters collected both quantitative metrics and qualitative comments from participants justifying their vulnerability detection decisions, resulting in a dataset of 263 free-text comments, each corresponding to a participant's rationale for detecting (or failing to detect) a vulnerability in a given code snippet. They were further annotated by other annotators.

\subsection{LLM Selection} \label{sec:sec:llm:selection} 

\begin{itemize}
    \item[M1:]  \textit{Models must be able to process spreadsheet files with annotations.} Annotations are supplied with several different codes at once, and the model must be able to mimic the use.
    \item[M2:] \textit{Different best models on LiveBench~\cite{white2024livebench}.} At the time of designing, the best-performing proprietary models for Reasoning Average score are ChatGPT-5 \cite{OpenAI_GPT5_2025}, o3 \cite{OpenAI_o3_2025} (Both from OpenAI), and Claude Sonnet 4 \cite{AnthropicClaude4_2025} (from Anthropic).
    \item[M3:] \textit{Open Source vs Proprietary Models.} Some of the models should be open source to provide a replicable experiment. The best-performing open-source models for Reasoning Average score are \modelthree \cite{DeepSeekV32_2025}, Kimi K2 Thinking \cite{team2025kimi} from Moonshot AI, and \modelfour \cite{yang2025qwen3} (from Alibaba).
    \item[M4:] \textit{Models must produce results of the expected size.} In preliminary experiments, some models failed to return the correct number of entries, either omitting part of the input or generating too many rows. Because this made it impossible to reliably match their outputs to the ground truth, we discarded models that did not produce the required number of rows.
\end{itemize}

For the proprietary models, we use the chat interface and the professional subscription, as we assume it is the best available model. We also investigated Grok \cite{xAI_Grok_2024} (from xAI), but it is not available in Italy at the time of designing. Besides \modelthree, Kimi K2 Thinking, and \modelfour, we also examined Mistral \cite{Mixtral_2024} and Llama \cite{LLaMA3_2024} through the Amazon Bedrock interface.

\subsection{Prompt Engineering}\label{sec:sec:prompting}

To answer \emph{RQ1}, we have crafted a series of prompts to enable the LLMs to function as an annotator. At the same time, to answer \emph{RQ2}, the prompts are applied \emph{sequentially}, involving the progressive enrichment of knowledge on code definitions, annotated examples, targeted clarifications, and interactive feedback mechanisms. 
\paragraph{Prompt Selection} Prompts are compared  to check whether increasing prompt complexity yields significant improvements (\emph{RQ2}): 
    \begin{itemize}
        \item[\textbf{\promptzero:} ]\textbf{The prompt contains only the task description and the name of the codes.} \promptzero describes research context, experiment design, spreadsheet structure (with code names), and annotation task, and it provides the file \emph{without} the applicable code of the other annotators. To generate a correct annotation without clues, the model can only rely on having seen something similar in the past.  

        \item[\textbf{\promptone:}] \textbf{The prompt includes the formal definition of the codes.} \promptone describes the expected behavior and provides an example for each column to be filled. This stage corresponds to the development of a formal codebook (Table~\ref{tab:input}).
        \item[\textbf{\prompttwo:}] \textbf{The prompt includes the examples of the other annotators.} \prompttwo provides the file \emph{with} the labels of other annotators on \emph{other comments}. Few-shot examples are considered a significant help for the LLMs, but they are not considered a good practice for human annotators, as they can be a source of bias and lack of independence. 
        \item[\textbf{\promptpilot:}] \textbf{The prompt asks for conflicting examples and provides clarifications from the researcher.}\promptpilot asks the pilot LLM to go through the human-annotated labels and extract two conflicting examples for each code, and ask what is unclear. A researcher then compiles the response to LLM's questions. This is done only once in the interaction.  
        \item[\textbf{\promptthree:}] \textbf{The prompt shows conflicting examples and the clarifications from researchers.} \promptthree shows LLM the conflicting examples obtained in the pilot interaction, as well as the researchers' answers to those ambiguities.
        \end{itemize}

The selection of prompts is based on a pilot experiment elaborated below, e.g., the prompt \promptthree is designed based on \promptpilot. 

\paragraph{Pilot Experiment}

\begin{figure}[t]
\centering
\begin{tabularx}{0.475\textwidth}{X}
\\
\begin{tcolorbox}[title=User  (\promptpilot), userbox]
\ldots 
Try to find two examples for each column that you find similar but have opposite labeling, so I can clarify subtle differences for you.
\end{tcolorbox}
\\
\begin{tcolorbox}[title=\pilotmodel, gptbox]
I’ve found contrastive examples from your annotations for \textit{each label}, where student responses are similar in phrasing or content but were marked differently. Below are \textit{two examples for each column}, showing the kind of subtle distinctions your reviewers made. You can now help me improve by clarifying these distinctions.

\textbf{Variable/Method identifiers are mentioned}:

-  `I think it had something to do with inv and debug' → marked 1

-  `I did not select anything because [...] I am not that experienced with C' → marked 0  

Clarify: Are only clear technical identifiers enough, or should \textit{any} code-ish word be counted?

\textbf{[Omit the contrastive examples for other codes \ldots] }

\end{tcolorbox}
\\
\begin{tcolorbox}[title=User (\promptpilot), userbox]

- \textbf{Variable/Method identifiers are mentioned}: The first sentence uses ``inv'' and ``debug'' in a way that you can understand they are not part of a natural language sentence, so they must be code identifiers. So the mark is 1. The second sentence is a natural language sentence, the only part for which you could have doubts is ``C'', but in this case it's part of the natural sentence expressing the student did not notice any vulnerability, maybe because he does not have enough experience in C programming

\textbf{[Omit the clarifications for other codes \ldots] }

\end{tcolorbox}
\caption{The pilot interaction where \promptpilot is constructed}\label{tab:p3pilot}

\end{tabularx}
\end{figure}

For the exploration of the initial design of the prompts, we employed OpenAI's \pilotmodel model~\cite{hurst2024gpt}  (interactive version) to conduct the pilot study for the prompts refinement. We chose \pilotmodel since this model is not used in the main experiments and can therefore reduce the bias. The output of each turn from the pilot LLM is manually reviewed for insights. Based on these outputs, we iteratively refined the input context and prompt content, incorporating targeted feedback and clarifications to address systematic errors or ambiguities observed in the model's predictions.

Prompts \promptzero\textasciitilde \promptthree were progressively specified and refined throughout the pilot experiment. In particular, during the specification of \promptthree the user first asked the pilot LLM to find contrastive examples from the provided samples, then provided clarification addressing those ambiguities.
Figure~\ref{tab:p3pilot} shows an illustrative example of \promptpilot, and Figure~\ref{tab:p3} shows the crafted \promptthree based on the pilot interaction.

The pilot LLM was asked to find contrastive examples. LLMs in the main experiment are not asked to label these contrastive examples to ensure that the corresponding answers are not inadvertently disclosed to the LLMs when providing prompt \promptthree.

%%%%%%%%%%%%%%
% \iffalse
\begin{figure}[t]
\centering
\begin{tabularx}{0.475\textwidth}{X} \\
\begin{tcolorbox}[title=User (\promptthree), userbox]

For each column, there are two examples that you might find similar but have opposite labeling. I can clarify the subtle differences for you.
 
\textbf{Variable/Method identifiers are mentioned}

-  `I think it had something to do with inv and debug' → marked 1

-  `I did not select anything because [...] I am not that experienced with C' → marked 0 

\emph{Clarify}: Are only clear technical identifiers enough, or should \textit{any} code-ish word be counted?

\emph{Clarification}: 
The first sentence uses inv'' and ``debug'' in a way that you can understand they are not part of a natural language sentence, so they must be code identifiers. So the mark is 1. [\ldots]

\textbf{[Omit the contrastive examples and the clarifications for other codes \ldots] }
\end{tcolorbox}
\end{tabularx}
\caption{The defined prompt \promptthree based on \promptpilot}\label{tab:p3}
\end{figure}
% \fi
%%%%%%%%%%%%%%%%%%

%All actual prompts we used in the experiments and the interaction in the pilot experiments will be available in the supplemental material.

% \begin{table*}[t]
% \caption{Iterations description}
% \longcaption{Description of the prompt(s) provided for each iteration. For each round, the model generated a new output attempt to complete the original task.}
% \begin{tabular}{lp{0.9\columnwidth}}
% \hline
% \textbf{\#} & \textbf{Description}                                                             \\ \hline
% \promptzero         & Described research context, experiment design, spreadsheet structure, task provide the file \emph{without} labels of the other annotators   \\                                     P1         & Described research context, experiment design, spreadsheet structure, task provide the file with labels of the other annotators \\ 
% P2         & Described in detail the expected behavior and provided an example for each column to be filled
% \\ 
% P3         & Asked GPT-4o to go through the human-filled rows, check if it would have marked the cells in the same way and otherwise try to understand and emulate the human reasoning                         \\ 
% P4-Pilot         & Asked the LLM one/off to go through the human-filled rows and extract two conflicting examples for each code asking what is unclear. A researcher compiled the response to all questions. \\
% P4 & Add all clarifications as the last prompt.\\
% \hline
% \end{tabular}
% \label{tab:iter}
% \end{table*}

\subsection{Formal Analysis}\label{sec:sec:analysis}
% \todo[inline]{MC: experiment. YG/FM: write and check.}
% \mc{We need to explain what is the dataset and how we use it again.}

% In this section, we define the analysis methods applied to the LLMs’ annotation results. First, we describe the traditional metrics used to initially analyze the raw output of LLMs' annotation (\S\ref{sec:analysis:metric}); next, we introduce the chance corrected accuracy which would be used as the primary evaluation metric (\S\ref{sec:analysis:HSS}); finally, statistical analysis are conducted to obtain results with reliability (\S\ref{sec:analysis:statistical}).

% \paragraph{Dataset and Notation}

% For the purpose of facilitating elaboration, the notations used for analysis are defined in Table~\ref{tab:notation}.
% \yg{not sure if needed}
% \begin{table}[th]
%     \centering
%     \caption{Notation Table}
%     \label{tab:notation}
%     \begin{tabular}{ll}
%     \hline
%     Notation& Description \\
%     \hline
%      a    &  \\
%      c    &  \\
%      l    &  \\
%      $\pi$    &  \\
%     \hline
%     \end{tabular}
% \end{table}

\paragraph{Success Metrics}\label{sec:analysis:metric}

For each run, across \TotalLLMs LLMs, \Totalprompts prompts, \Totalcodes codes, and \Totalannotators annotators, we obtain a tuple which summarizes the result of the annotation of the \Totalcomments comments.

\noindent $\langle \ell, \prompt , \code, \annotator, TP_{\ell \prompt  \code \annotator}, FP_{\ell \prompt  \code \annotator}, , FN_{\ell \prompt \code \annotator}, FN_{\ell \prompt \code \annotator}\rangle$, where: 

\begin{itemize}
    \item $TP_{\ell \prompt \code \annotator}$ is the number of comments where $\code=1$ by the annotator $a$ and $\code=1$ by the LLM $\ell$ with prompt \prompt; 
    \item $FP_{\ell \prompt \code \annotator}$ is the number of comments where $\code=0$ by the annotator \annotator and $\code=1$ by the LLM $\ell$ with prompt \prompt;
    \item $FN_{\ell \prompt \code \annotator}$ is the number of comments where $\code=1$ by the annotator \annotator and $\code=0$ by the LLM $\ell$ with prompt \prompt;
    \item $TN_{\ell \prompt \code \annotator}$ is the number of comments where $\code=0$ by the annotator \annotator and $\code=0$ by the LLM $\ell$ with prompt \prompt.   
\end{itemize}

We define by $correct_{\ell,  \prompt, \code, \annotator}=TP_{\ell, \prompt, \code, \annotator}+TN_{\ell, \prompt, \code, \annotator}$ the number of correctly coded comment by the LLM $\ell$ for code \code, 
we denote by $pos_{\ell, \prompt, \code, \annotator}=TP_{\ell, \prompt, \code \annotator}+FN_{\ell, \prompt, \code, \annotator}$ the number of positives according to the human annotator, and by $alert_{\ell, \prompt, \code, \annotator}=TP_{\ell, \prompt, \code, \annotator}+FP_{\ell, \prompt, \code, \annotator}$ we denote the number of comments marked as 1 for code \code. %annotated by LLM $\ell$.

\paragraph{Cohen's Kappa $\kappa$ and Chance Corrected Accuracy}
As mentioned in the introduction, the first step is to make sure that we do not hold LLMs to higher standards than human annotators. Cohen's Kappa is a robust indicator of agreement in which the percentage of codes in agreement ($TP$ and $TN$) between two coders (in our case, the LLM and one of the annotators) is normalized by the percentage of the agreement that could have been obtained by chance. 
For the sake of simplicity, we remove in the subsequent equations the indexes describing the LLM, the column, and the annotator.

\begin{eqnarray}
E[Agree] & = &  \frac{alert}{N}\cdot \frac{pos}{N} + \left(1-\frac{alert}{N}\right)\cdot \left(1-\frac{pos}{N}\right)\\
Agree (=Acc) & = &  \frac{correct}{N} \label{eq:agree}
    % E[Acc] & = &  \frac{TP+FP}{N}\cdot \frac{TP+FN}{N} + \nonumber \\
    % & & + \left(1-\frac{TP+FP}{N}\right) \left(1-\frac{TP+FN}{N}\right) \\
    \\
    \kappa (=Acc^*) & = & \frac{Agree-E[Agree]}{1-E[Agree]} \label{eq:kappa}
\end{eqnarray}

The definition of Agreement (\ref{eq:agree}) is actually \emph{identical to accuracy if we consider each annotator as the ground truth}, and Cohen's Kappa (\ref{eq:kappa}) is actually identical to the definition of \emph{chance corrected accuracy}. This indicator is quite common in the weather prediction literature \cite{barnston1992correspondence}: 
In the presence of an unbalanced dataset, some of the performance metrics might be obtained by pure chance, just because we have too many (or too few) events of interest in the system. To avoid such mistakes, we need to compute the expected values of true positives or true negatives that we would have obtained by chance and subtract them from the values we obtain with the experiment. The chance corrected accuracy is also known in the weather forecasting literature as the Heidke skill index (\emph{HSS}). 

Common guidelines \cite{warrens2015five} state that an interval of $[0.0,0.2]$ indicates slight agreement, $(0.2, 0.4]$ fair agreements, $(0.4,0.6]$ moderate agreement, $(0.6,0.8]$ substantial agreement, and 
$(0.8,1.0]$ indicates almost perfect agreement. We can therefore use the same guidelines to compare chance-corrected accuracy against the bands of Cohen's $\kappa$. We also have to account for the possibility that chance-corrected accuracy is negative, namely that the LLM is actually worse than randomly guessing human annotations. 
We summarize these levels in Table \ref{tab:thresholds} and adapt it to the LLMs.

\begin{table}[t]
    \caption{Proposed Thresholds for Human Replaceability}
    \label{tab:thresholds}
    \centering
    \begin{tabular}{cp{0.8\columnwidth}}
    \hline
    $\kappa (Acc^*)$ &  Replaceability and Rationale\\
     \hline
    $<0$   & \textbf{Hallucinate} (\StarHallucinate) - The LLM has worse performance than random guessing in controlled experiments. It is \emph{not} recommended to use it. \\
    %$[0.0,0.2]$  &
    $[0.0,0.2]$ & \textbf{Unusable} (\StarUnusable) - The LLM will rarely generate the expected annotations and will almost always disagree with the human annotator.\\
    %$(0.2, 0.4]$&
    $(0.2, 0.4]$ & \textbf{Limited} (\StarLimited), The LLM only partly generates the annotations of a human coder. It might not converge when resolving conflicts with other annotators.\\
    %$(0.4,0.6]$ &
    $(0.4,0.6]$ & \textbf{Moderate} (\StarModerate), The LLM is not able to generate human annotation for around half the time. It might or might not converge when discussing conflicts.\\
    %$(0.6,0.8]$&
    $(0.6,0.8]$ & \textbf{Substantial} (\StarSubstantial) - The LLM will generate codes close to a human annotator in practice. Multiple interactions over conflicts are likely to converge.\\
    %$(0.8,1.0]$&
    $(0.8,1.0]$ & \textbf{Almost Perfect} (\StarAlmostPerfect) - The LLM would generate almost the same annotations as human coders, or at least will disagree with them only in a few cases.\\
\hline
    \end{tabular}
\end{table}

Since we have multiple annotators, when reporting a value, we take the average across annotators.

\paragraph{Analysis and Statistical Tests}\label{sec:analysis:statistical}

To answer \emph{RQ1}, we first compute each metric and we report the average of the results across $A$ annotators. Since an LLM might be better suited to some types of codes than others (e.g., better at understanding the presence of program identifiers than security keywords), we also consider the individual coders as a second detailed comparison.

Then compare the relative effectiveness of LLM $\ell_i$ globally, we perform a Wilcoxon paired test as a statistical test to compare the set of kappa's value $\kappa_{\ell, \prompt, \code,\annotator}$ for each prompt, considering each code and annotator as pairs.
\begin{align}
    W\left(\left\{\left.\left\langle \kappa_{\ell_1, \prompt, \code, \annotator} ~,~ \kappa_{\ell_2,\prompt, \code, \annotator}\right\rangle\right | \prompt, \code, \annotator\right\}\right)  \mbox{ for } \ell_1  \mbox{ and } \ell_2\label{eq:wilcoxon:llm}
\end{align}

% Since we are running multiple comparisons, we consider significant only the tests with a p-values $p\leq \frac{5\%}{P}$.

To answer \emph{RQ2}, first, we evaluate the effectiveness of the prompts irrespective of the LLMs. We aggregate the results over the codes and annotators for each LLM $\ell$ and prompt $\prompt$. We perform the pairwise Wilcoxon test to compare the prompts with increasing level of effort, i.e., comparing prompt $\prompt _{i-1}$ with prompt $\prompt _{i}$.
\begin{align}
    W\left(\left\{\left.\left\langle \kappa_{\ell, \prompt_{i-1}, \code, \annotator}, \kappa_{\ell,\prompt_i, \code, \annotator}\right\rangle\right | \ell, \code, \annotator\right\}\right) \label{eq:wilcoxon:prompts}
\end{align}
 Since prompts are compared in increasing order ($\prompt _{i-1}$ is compared with $\prompt _{i}$, which is then compared with $\prompt_{i+1}$, etc. they are independent tests, and we do not need to perform a Bonferroni correction.

We can refine the analysis by comparing each LLM while keeping the prompt constant, essentially repeating the test in Equation (\ref{eq:wilcoxon:llm}) for all prompts. In this setting, we need to perform a Bonferroni correction by considering as statistically significant only those tests with a p-value $p\leq \frac{5\%}{\TotalLLMs}$ where \TotalLLMs is the number of LLMs compared because we are repeating the previous test again now with a different metric each time. 

% \begin{table*}[t]
%     \caption{Proposed Thresholds for Human Replaceability}
%     \label{tab:thresholds}
%     \centering
%     \begin{tabular}{p{0.2\columnwidth}p{0.2\columnwidth}p{0.3\columnwidth}p{1.1\columnwidth}}
%     \hline
%     $\kappa$& $ACC^*$ &  Replaceability Level  & Rationale\\
%      \hline
%     -&$<0$   & failure & The LLM has worse performance than random guessing in controlled experiments. It is \emph{not} recommended to use it. \\
%     $[0.0,0.2]$  &$[0.0,0.2]$ & unfeasible &  The LLM will rarely generate the expected annotations and will almost always disagree with the human annotator.\\
%     $(0.2, 0.4]$&$(0.2, 0.4]$ & limited & The LLM only partly (way less than half the times) generates the annotations of a human coder. It might not converge when resolving conflicts with other annotators.\\
%     $(0.4,0.6]$ &$(0.4,0.6]$ & moderate & The LLM is not able to generate human annotation for around half the times. It might or might not converge when discussing conflicts.\\
%     $(0.6,0.8]$&$(0.6,0.8]$ & substantial & The LLM will generate codes close to a human annotator in practice. Multiple interactions over conflicts are likely to converge.\\
%     $(0.8,1.0]$&$(0.8,1.0]$ & almost perfect & The LLM would generate almost the same annotations of human coders (or at least will disagree with them only in few cases)\\
% \hline
%     \end{tabular}
% \end{table*}

\section{Results} \label{sec:results}

\subsection{RQ1: LLMs Ability on Automatic Annotation}
Table~\ref{tab:globalresults} reports the summary with the average precision, recall, accuracy, and Cohen's Kappa (chance-corrected accuracy) for our experiments where we tried to replace a human annotator with an LLM for the best prompt \promptthree that we have presented in \S\ref{sec:sec:prompting}. Values are averaged across annotators and columns. e.g. $\hat{\kappa}_{\ell}=\frac{1}{\Totalannotators}\frac{1}{\Totalcodes}\sum_{\annotator,\code}\kappa_{\ell, \prompt_{3+},  \code, \annotator}$ to provide a first global assessment.
As we can see from the Table the result is negative. We cannot reliably assume that an LLM can \emph{generally} replace a human annotator.

\begin{table}[t]
\centering
\caption{Average Results with the Final Prompt}
\label{tab:globalresults}
% \todo[inline]{MC: final prompt results}
\begin{minipage}{0.95\linewidth}
Based on the qualitative levels from Table~\ref{tab:thresholds} for Cohen's $\kappa$ guidelines \cite{warrens2015five}, we see that LLMs cannot reliably replace human annotators. We also see the importance of using robust metrics not affected by unbalanced datasets.
\end{minipage}
\begin{tabular}{lccccl}
    \hline
     LLM & Precision & Recall & Acc & $\kappa/Acc^*$ & Replace \\
     % LLM & Precision & Recall & Acc & Acc$^*$ & Replaceability\footnotemark[1] \\
    \hline
    \gptmodel & 0.62 & 0.47 & 0.74 & 0.26 & \StarLimited  \\
    \claudemodel & 0.57 & 0.51 & 0.74 & 0.30 & \StarLimited  \\
    \modelthree  & 0.79 & 0.64 & 0.83 & 0.53  &\StarModerate \\
    \modelfour & 0.74 & 0.73 & 0.86 & 0.61 &\StarSubstantial \\
    \hline
\end{tabular}
\begin{footnotesize}Table~\ref{tab:thresholds}: Substantial - \StarSubstantial, Moderate -\StarModerate, Limited - \StarLimited.
\end{footnotesize}
% \caption{Global Results with the final prompt}
% \label{tab:globalresults}
\end{table}

% The results of Wilcoxon two-sided test between \gptmodel and \claudemodel is $W = 20,~p=0.82$ so we cannot conclude that either model is superior to the other model. \todo[inline]{FM: Add here the text of the other models.}

Human annotators are also error-prone and might disagree with each other. Therefore, a too high bar might not correspond to practical use. The next step investigate whether the poor (overall) performance is only due to some specific codes that are also hard for human annotators.

Table~\ref{tab:globalresultscode} shows the summary of the results for the chance-corrected accuracy of each code for the final prompt. 
% \todo[inline]{FM: I d do not remember how do you calculate this one so, I put this text below. If we give them the entire thing then delete it. YG: yes that was the decision}
In this set-up we have given LLMs an \emph{easier task} as we only consider the comments on which the human annotators agree with the reviewers. So LLMs were given codes where there is 100\% agreement with humans. Once again, we only consider our best prompt \promptthree.

\begin{table*}[ht]
\centering
\caption{The Cohen's $\kappa$ of the human annotator/LLM's agreement with the reviewers across codes}\label{tab:globalresultscode}
\begin{minipage}{0.95\textwidth}
%\footnotesize 
%For readability We only mark the cases in which the replaceability is at least moderate. 
In spite of giving the LLM an easy task (by supplying it only the comments on which there was almost 100\% agreement among the humans), it could not successfully annotate most codes. 
Only on comments that describe the absence of a vulnerability (\texttt{NoVul}) \modelthree and \modelfour are able to get close to the human agreement. \gptmodel and \claudemodel exhibit very low performance on certain columns, but other models also have column-specific limitations.
\end{minipage}

% \begin{tabular}{lccccc}
%\begin{tabular}{lccccccccccccccc}
% Code &Human  &\gptmodel&\claudemodel&\modelthree&\modelfour\\
%& \multicolumn{3}{c}{Human} & \multicolumn{3}{c}{\gptmodel}&\multicolumn{3}{c}{\claudemodel}&\multicolumn{3}{c}{\modelthree}&\multicolumn{3}{c}{\modelfour}\\
%Code & Mean & Min & Mean & Max& Min & Mean & Max& Min & Mean & Max& Min & Mean & Max \\
% \usepackage[normalem]{ulem}

\begin{tabular}{l@{\hspace{0.5cm}}rl@{\hspace{0.8cm}}rl@{\hspace{0.8cm}}rl@{\hspace{0.8cm}}rl@{\hspace{0.8cm}}rl}
\hline
      &                & \hspace{-0.5cm} Human* &                & \hspace{-0.5cm}\gptmodel &                & \hspace{-0.5cm}\claudemodel &                & \hspace{-0.5cm}\modelthree &                & \hspace{-0.5cm}\modelfour \\
      & $\hat{\kappa}$ & Agreement                               & $\hat{\kappa}$ & Replace                                                  & $\hat{\kappa}$ & Replace                                                     & $\hat{\kappa}$ & Replace                                                    & $\hat{\kappa}$ & Replace                                                   \\ \hline
Var   & 0.98           & \StarAlmostPerfect       & 0.00           & \StarUnusable                             & 0.52           & \StarModerate                                & 0.79           & \StarSubstantial                            & 0.79           & \StarSubstantial                           \\
Lin   & 0.94           & \StarAlmostPerfect       & 0.57           & \StarModerate                             & 0.46           & \StarModerate                                & 0.31           & \StarLimited                                & 0.42           & \StarModerate                              \\
Key   & 0.98           & \StarAlmostPerfect       & 0.30           & \StarLimited                              & 0.14           & \StarUnusable                                & 0.34           & \StarLimited                                & 0.56           & \StarModerate                              \\
Vul   & 0.98           & \StarAlmostPerfect       & 0.50           & \StarModerate                             & 0.44           & \StarModerate                                & 0.39           & \StarLimited                                & 0.50           & \StarModerate                              \\
Sec   & 0.98           & \StarAlmostPerfect       & 0.11           & \StarUnusable                             & 0.15           & \StarUnusable                                & 0.50           & \StarModerate                               & 0.66           & \StarSubstantial                           \\
Exp   & 0.98           & \StarAlmostPerfect       & 0.12           & \StarUnusable                             & 0.20           & \StarUnusable                                & 0.56           & \StarModerate                               & 0.64           & \StarSubstantial                           \\
NoVul & 0.98           & \StarAlmostPerfect       & 0.30           & \StarLimited                              & 0.42           & \StarModerate                                & 0.82           & \StarAlmostPerfect                          & 0.85           & \StarAlmostPerfect                         \\
Unsr  & 0.93           & \StarAlmostPerfect       & 0.38           & \StarLimited                              & 0.16           & \StarUnusable                                & 0.66           & \StarSubstantial                            & 0.59           & \StarModerate                              \\
Unclr & 0.95           & \StarAlmostPerfect       & 0.00           & \StarUnusable                             & 0.13           & \StarUnusable                                & 0.44           & \StarModerate                               & 0.41           & \StarLimited                               \\ \hline
Total & 0.97           & \StarAlmostPerfect       & 0.25           & \StarLimited                              & 0.29           & \StarLimited                                 & 0.54           & \StarModerate                               & 0.60           & \StarModerate                              \\ \hline
\end{tabular}

\begin{footnotesize}Table~\ref{tab:thresholds}: Almost Perfect - \StarAlmostPerfect, Substantial - \StarSubstantial, Moderate -\StarModerate, Limited - \StarLimited, Unusable - \StarUnusable, Hallucinate - \StarHallucinate).
\end{footnotesize}
\end{table*}

As we can see, on individual codes, the results are partly consistent across models with some notable exceptions. All models perform very poorly on some codes, such as identifying the unclear answer (\texttt{Unclr}). They have a better success on spotting the negative answer `I don’t find vulnerability' (\texttt{Vul}) and to identify whether there are Variable/Method identifiers mentioned (\texttt{Var}).
% Only on the identification of code identifiers \claudemodel achieves a moderate success while \gptmodel is essentially random guessing. On the presence of line numbers in the code \gptmodel achieves a moderate success while \claudemodel is still below the 40\% threshold.

\begin{tcolorbox}[colback=gray!3, colframe=black!30, boxrule=0.4pt, arc=1mm]
\textbf{Negative Answer to RQ1:} The tested LLMs are not generally able to match the human annotators except for some specific code and only for some specific models. Surprisingly, open source models perform generally better than proprietary LLMs. LLMs are not yet ready to replace human annotators in security-relevant comments.
\end{tcolorbox}

\subsection{RQ2: Impact of Prompt Efforts}

Table \ref{tab:global:results:prompt} shows the results per prompt across the four LLMs as we are interested in determining whether there is a significant difference between the prompts, irrespective of the LLMs.
\begin{table}[ht]
\centering
\caption{Cohen's $\kappa$ (Chance Corrected Accuracy) per Prompt}
\label{tab:global:results:prompt}
\begin{minipage}{0.95\linewidth}
The only significant difference is obtained by providing the LLMs with the definition of the codes (from \promptzero to \promptone). The large effort to further improve the prompts does not bring significant gain across different LLMs and also seems to confuse the model.
\end{minipage}
\begin{tabular}{lccccl}
    \hline
     Prompt & Precision & Recall & Acc & $\kappa (Acc^*)$ & Replaceability\\
    \hline
    \emph{\promptzero} & 0.63 & 0.51 & 0.78 & 0.37 & \StarLimited  \\
    \emph{\promptone} & 0.67 & 0.56 & 0.79 & 0.42 & \StarModerate\\
    \emph{\prompttwo} & 0.59 & 0.54 & 0.76 & 0.35 & \StarLimited \\
    \emph{\promptthree} & 0.68 & 0.58 & 0.79 & 0.42 & \StarModerate \\
    \hline
\end{tabular}
\begin{footnotesize}Table~\ref{tab:thresholds}: Substantial - \StarSubstantial, Moderate -\StarModerate . \end{footnotesize}
% \footnotetext[1]{Qualitative levels are defined in Table \ref{tab:thresholds} similarly to the Cohen's $\kappa$ guidelines \cite{warrens2015five}.}
\end{table}
Once again, we do not see a major difference. After the first prompt, where the chance-corrected accuracy moves from 37\% to 42\%, the values linger in the area. Refining the prompts does not yield major changes. 

Table \ref{tab:wilcoxon:prompts} reports the results of Wilcoxon paired tests between increasingly sophisticated prompts. While the comparison between \promptzero and \promptone yields a $p$-value lower than 5\%, such a value is not statistically significant after the Bonferroni correction, where we divide by the number of multiple comparisons $p >0.05/(P-1)=0.017$ where $P$ is the number of compared prompts in order of sophistication. 
\begin{table}
\centering
\caption{Wilcoxon paired tests between increasingly sophisticated prompts}\label{tab:wilcoxon:prompts}
\begin{tabular}{cccc}
\hline
 \multicolumn{2}{c}{Contrast}& Wilcoxon's $W$
& $p$-value  \\
\hline
 \emph{\promptzero} & \emph{\promptone}&181&%\cellcolor{green!25}
 0.037\\
\emph{\promptone} &\emph{\prompttwo}&292&0.890\\
\emph{\prompttwo} &\emph{\promptthree}&30&\textbf{0.004}\\
\hline
\end{tabular}
\end{table}
So we cannot really conclude that spending time refining the prompts yields any better outcome by the LLMs.
This finding persists even if separating the results by model, see Table \ref{tab:wilcoxon:all}.
\begin{table}[ht]
\centering
\caption{Wilcoxon paired tests across LLMs and prompts}\label{tab:wilcoxon:all}
\begin{tabular}{c|cccc}
\hline
LLM & \multicolumn{2}{c}{Contrast}& Wilcoxon's $W$
& $p$-value  \\
\hline
\multirow{3}{*}{\gptmodel}  & \emph{\promptzero} & \emph{\promptone}&3&0.04\\
 & \emph{\promptone} &\emph{\prompttwo}&45&1.0\\
 &\emph{\prompttwo} &\emph{\promptthree}&0&\textbf{0.004}\\ 
 \hline
\multirow{3}{*}{\claudemodel}   & \emph{\promptzero} & \emph{\promptone}&9&0.06\\
 & \emph{\promptone} &\emph{\prompttwo}&31&0.85\\
 &\emph{\prompttwo} &\emph{\promptthree}&16&0.42\\
\hline
\multirow{3}{*}{\modelthree}   & \emph{\promptzero} & \emph{\promptone}&12&0.23\\
 & \emph{\promptone} &\emph{\prompttwo}&3&0.63\\
 &\emph{\prompttwo} &\emph{\promptthree}&0&0.5\\
\hline
\multirow{3}{*}{\modelfour}   & \emph{\promptzero} & \emph{\promptone}& 28&0.75\\
 & \emph{\promptone} &\emph{\prompttwo}&0&\textbf{0.001} \\
 &\emph{\prompttwo} &\emph{\promptthree}&3&1.0\\
\hline
%\multicolumn{5}{p{0.8\textwidth}}{\footnotesize Cells with highlighted background indicates statistical significance, and underlining denotes significance after Bonferroni correction. }\\
\end{tabular}
\end{table}

\begin{figure*}[t]
    \centering
    % \begin{subfigure}{0.41\linewidth}
    %     \centering
    %     \includegraphics[width=\linewidth]{HSS_effort.png}
    %     \subcaption{exp1, graphic of \gptmodel}
    % \end{subfigure}
    % \hspace{0.05\linewidth}
    % \begin{subfigure}{0.41\linewidth}
    %     \centering
    %     \includegraphics[width=\linewidth]{HSS_effort.png}
    %     \subcaption{exp1, graphic of \claudemodel}
    % \end{subfigure}
    \begin{subfigure}{0.23\linewidth}
        \centering
        \includegraphics[width=\linewidth]{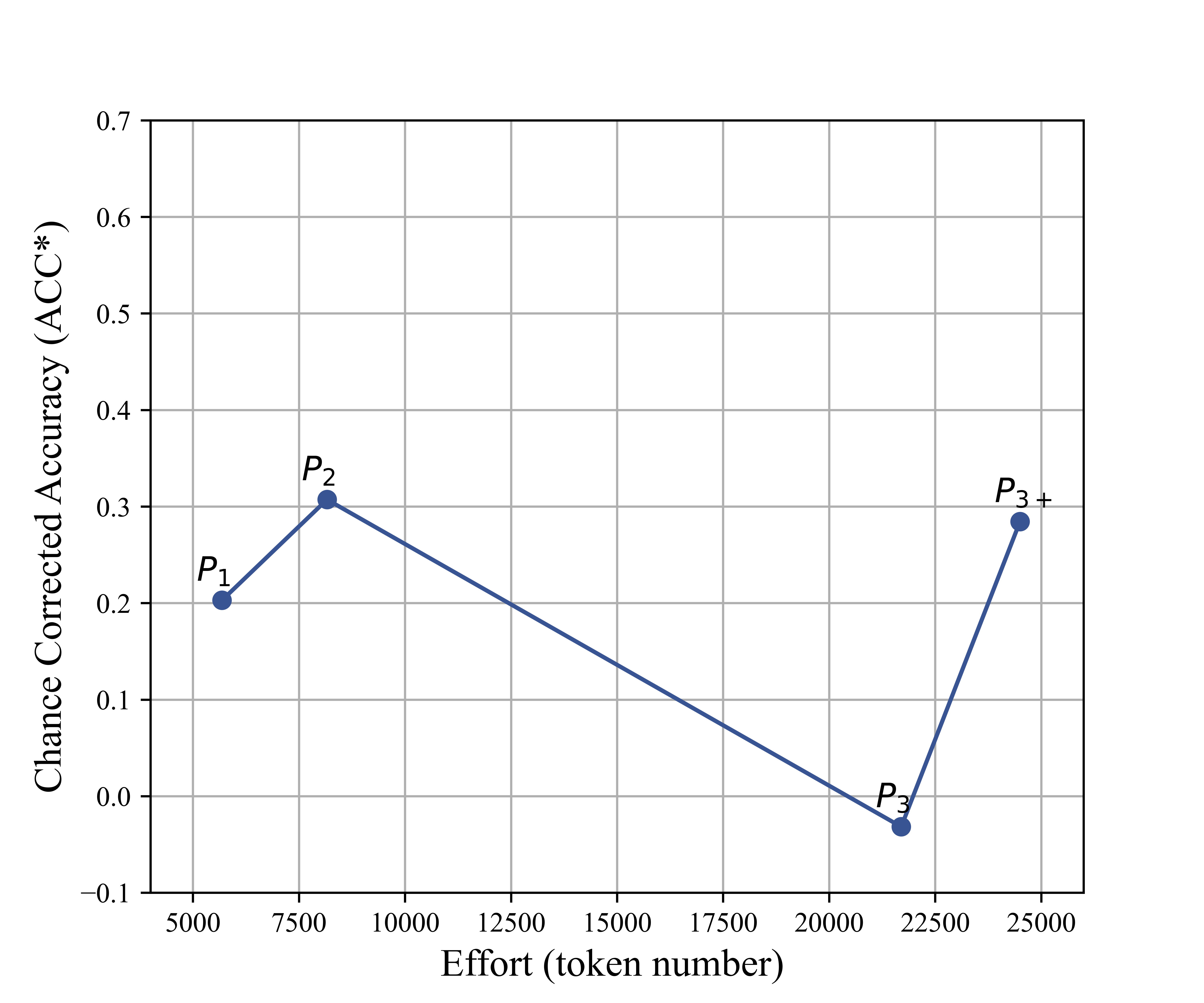}
        \subcaption{\gptmodel}
    \end{subfigure}
    % \hspace{0.05\linewidth}
    \begin{subfigure}{0.23\linewidth}
        \centering
        \includegraphics[width=\linewidth]{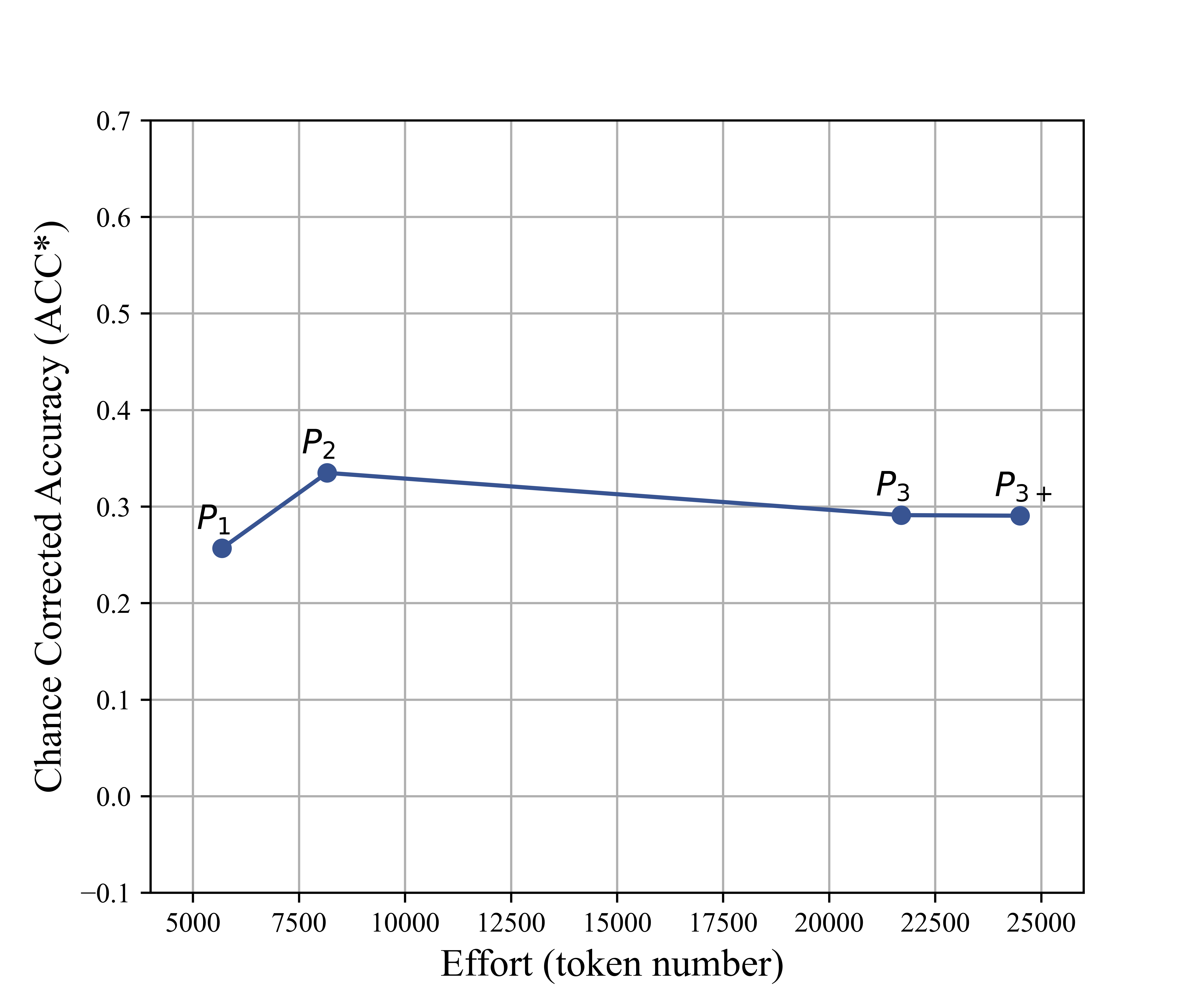}
        \subcaption{\claudemodel}
    \end{subfigure}
    \begin{subfigure}{0.23\linewidth}
        \centering
        \includegraphics[width=\linewidth]{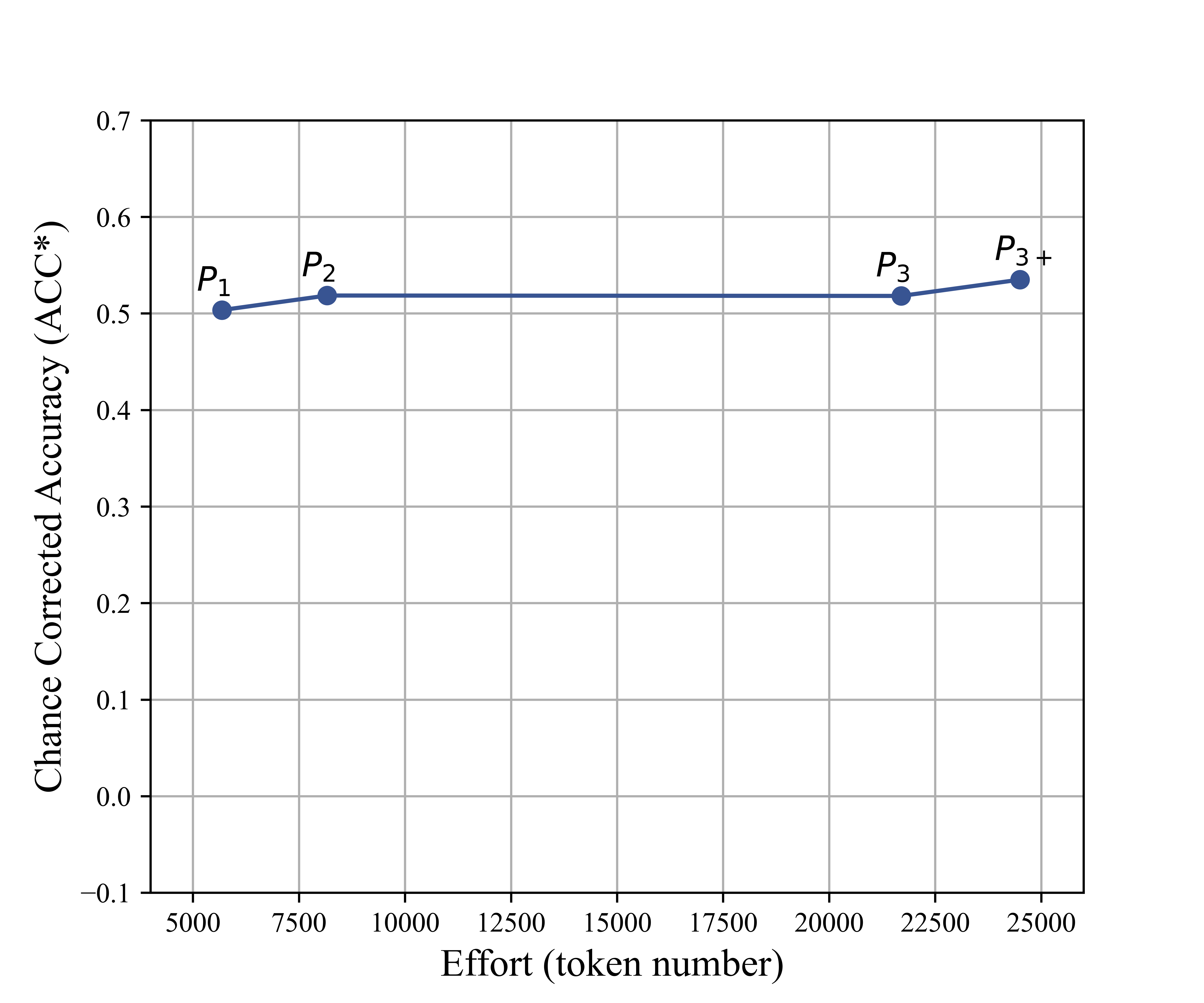}
        \subcaption{\modelthree}
    \end{subfigure}
    % \hspace{0.05\linewidth}
    \begin{subfigure}{0.23\linewidth}
        \centering
        \includegraphics[width=\linewidth]{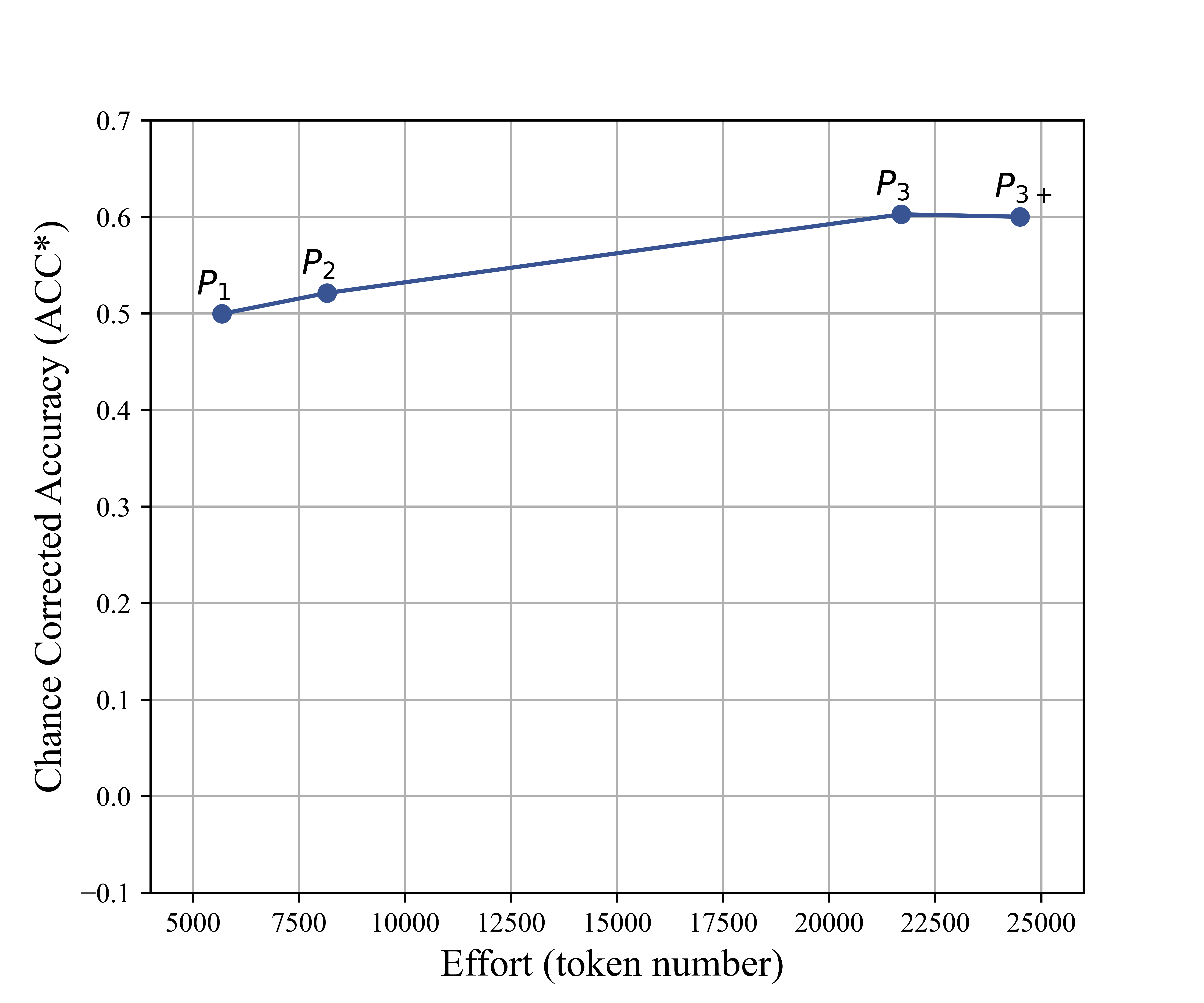}
        \subcaption{\modelfour}
    \end{subfigure}
    % \longcaption{After the first iteration, no significant improvements in chance corrected accuracy were observed, regardless of the invested effort.}
    \caption{Average Cohen's $\kappa$ (Chance Corrected Accuracy if human were ground truth) vs Effort (Size of Prompts)}\label{fig:HSS:effort}
\end{figure*}

\begin{figure*}[t]
    \centering
    \begin{subfigure}{0.23\linewidth}
        \centering
        \includegraphics[width=\linewidth]{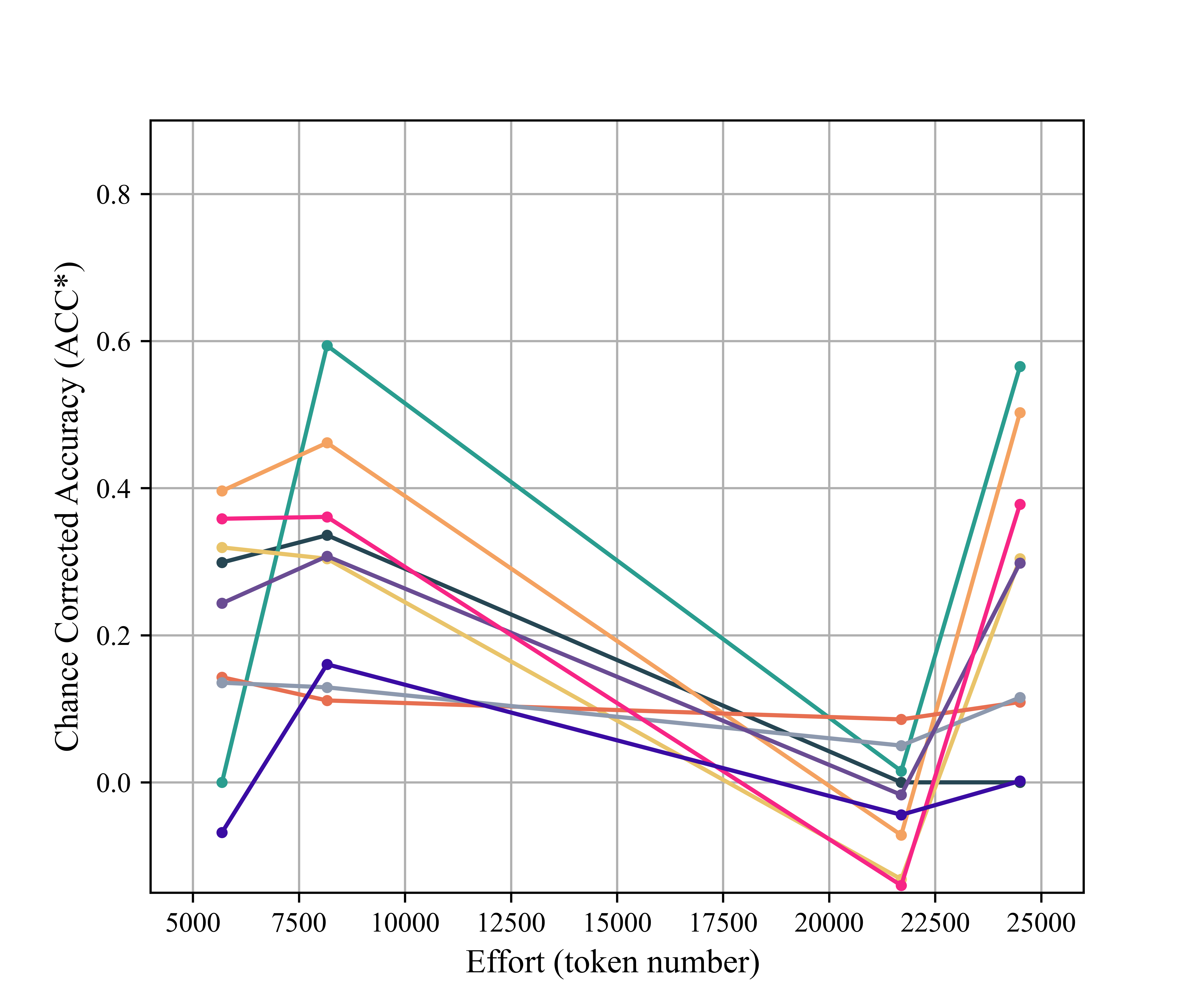}
        \subcaption{\gptmodel}
    \end{subfigure}
    % \hspace{0.2\linewidth}
    \begin{subfigure}{0.23\linewidth}
        \centering
        \includegraphics[width=\linewidth]{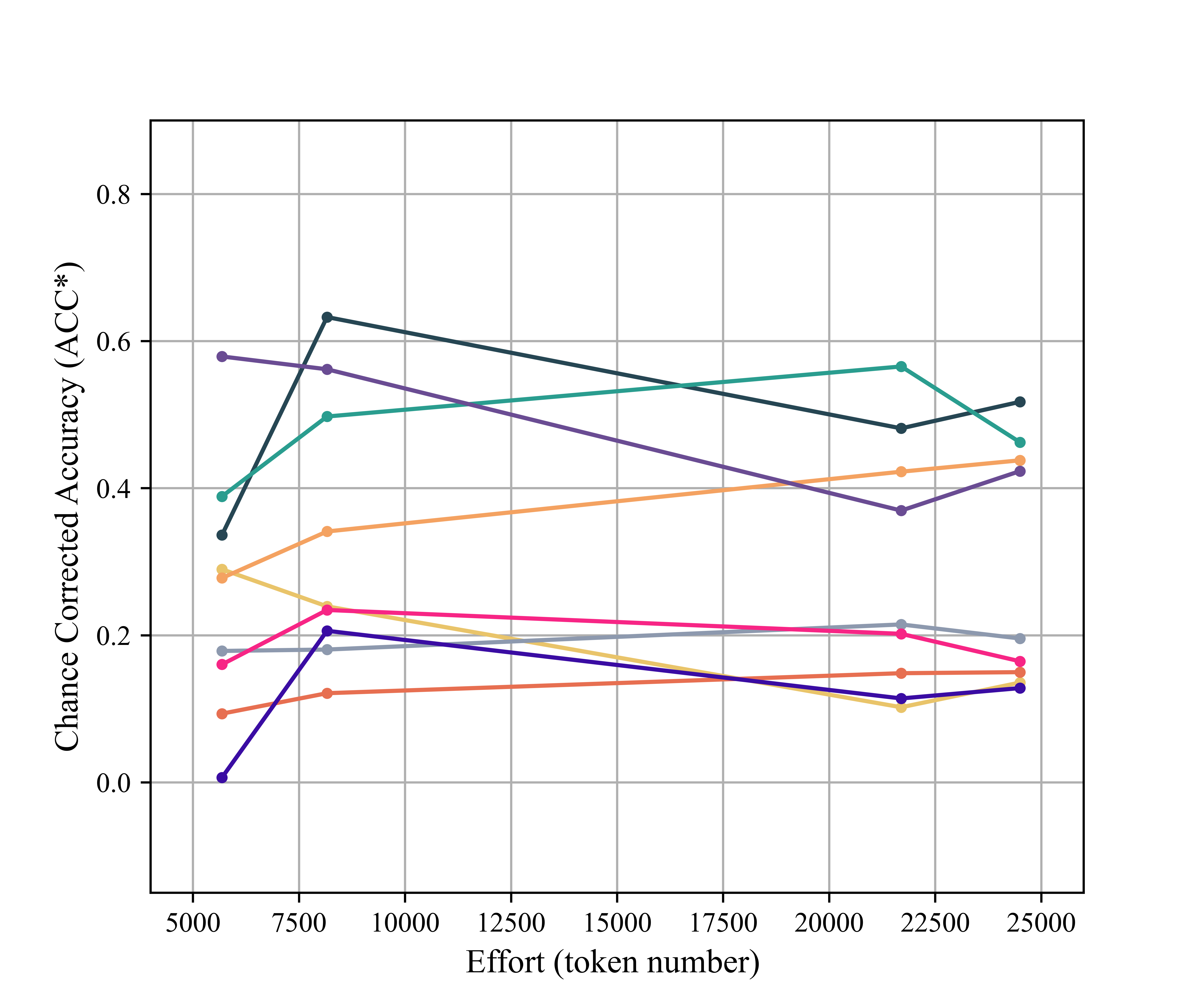}
        \subcaption{\claudemodel}
    \end{subfigure}
    \begin{subfigure}{0.23\linewidth}
    \centering
    \includegraphics[width=\linewidth]{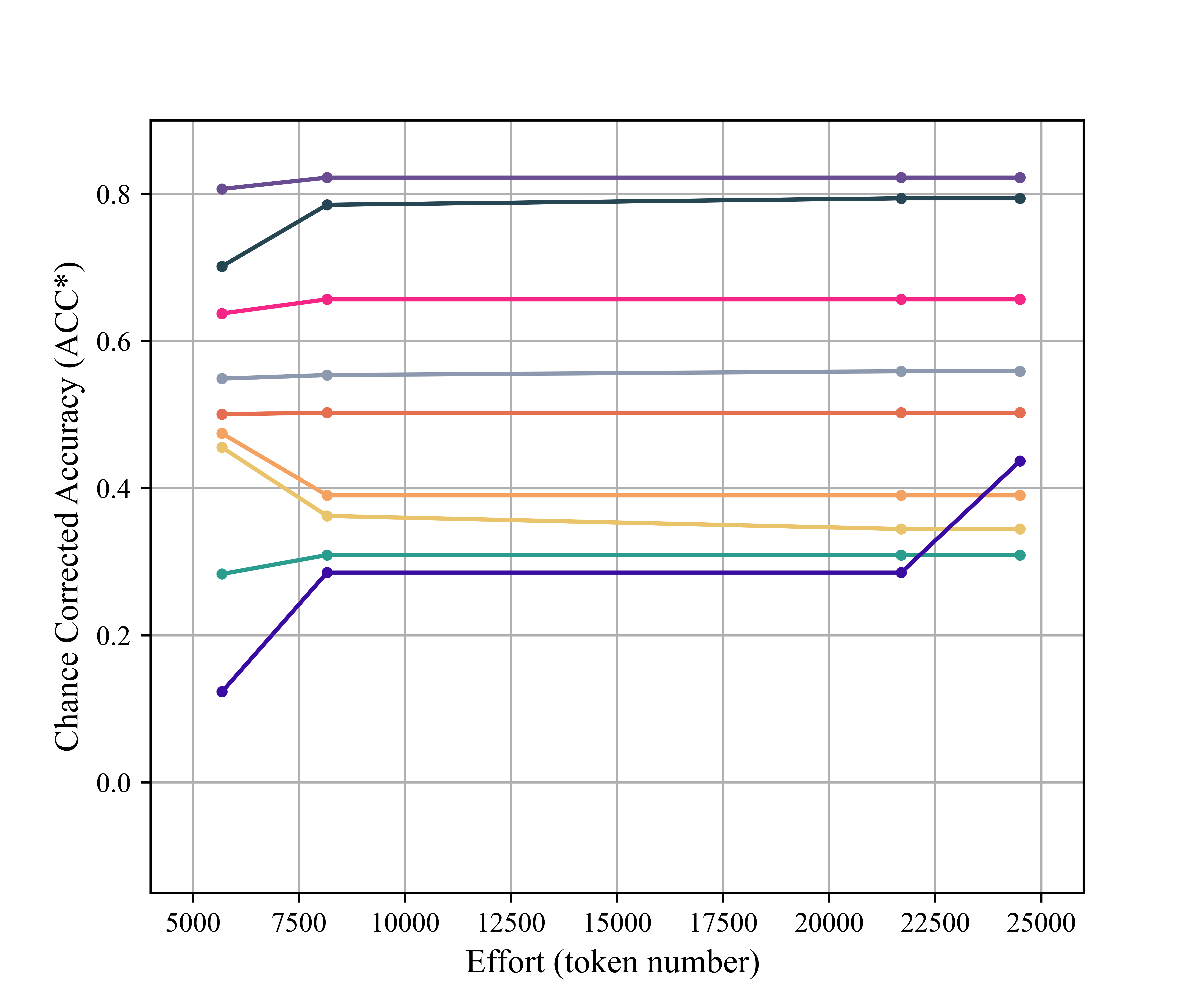}
    \subcaption{\modelthree}
    \end{subfigure}
    % \hspace{0.2\linewidth}
    \begin{subfigure}{0.23\linewidth}
    \centering
    \includegraphics[width=\linewidth]{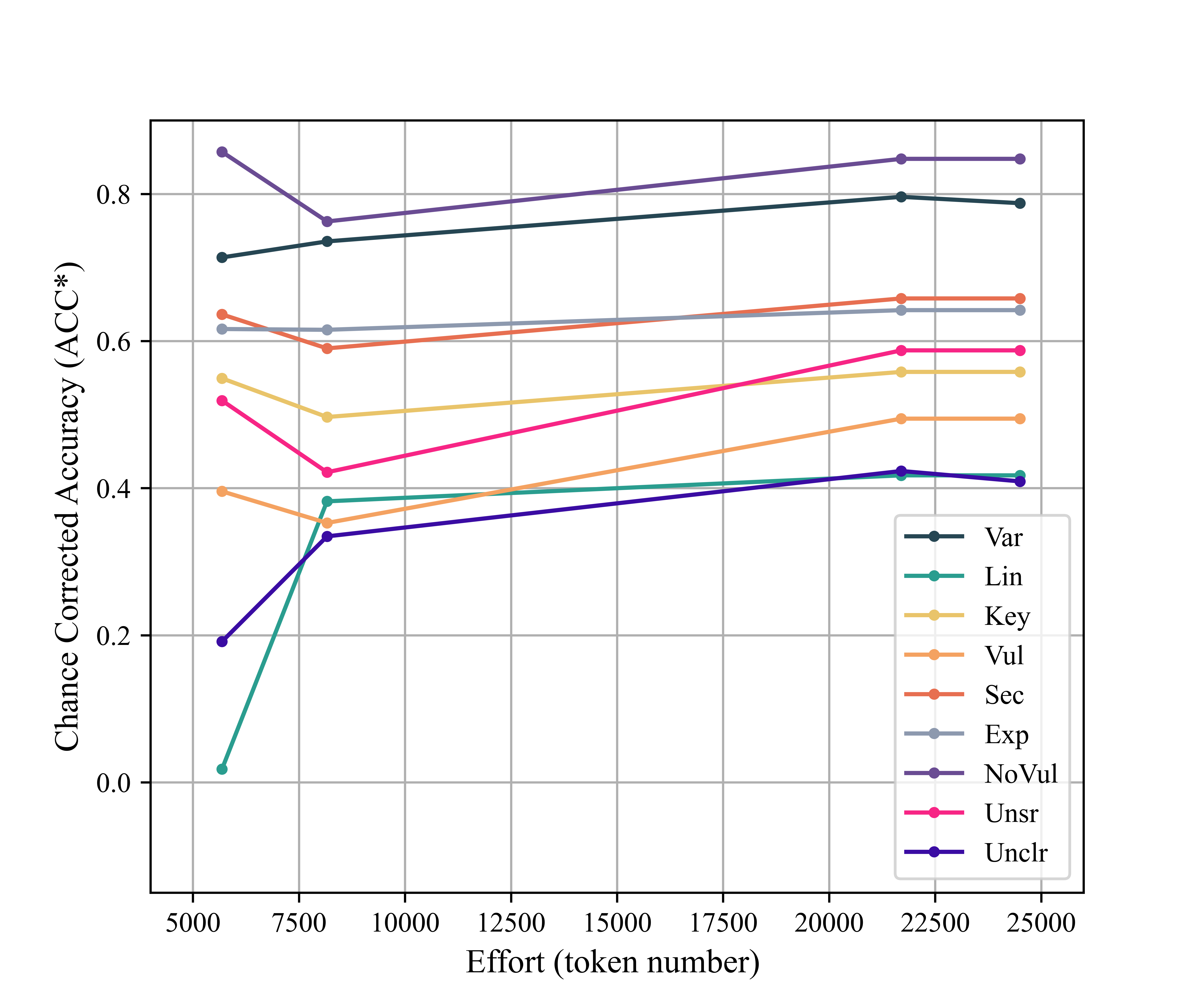}
    \subcaption{\modelfour}
    \end{subfigure}
    \caption{Chance corrected accuracy against the required effort in tokens, with each curve corresponding to a different code.}\label{fig:HSS:codes}
\end{figure*}

These findings are particularly interesting as the different prompts have different budgets in terms of tokens. Figure \ref{fig:HSS:effort} shows the results in terms of chance-corrected accuracy against the tokens used to obtain those results. Since the models have different ways to count tokens, and it is unclear how the Excel files are precisely counted, we opted for a third-party software (AWS Bedrock), which can be used to query several models to obtain a token count. Figure \ref{fig:HSS:codes} shows the results for each code.

\begin{tcolorbox}[colback=gray!3, colframe=black!30, boxrule=0.4pt, arc=1mm]
\textbf{Negative Answer to RQ2:} Prompt refinements with increasing efforts in terms of tokens do not uniformly guarantee 
a good performance across LLMs for security-specific annotations. There is only a significant gain at the beginning when a formal definition of the codes is provided with the positive examples. Even proprietary models might be significantly confused by moire data.
\end{tcolorbox}

\subsection{Qualitative Analysis}

\paragraph{LLMs’ elaboration of tabular data.}\label{sec:llmsXtabulardata}
Several LLMs initially considered were excluded due to technical limitations encountered during experimentation. We first evaluated models via the Amazon Bedrock~\cite{bhattacharjee2025introduction} platform. Although Claude~4 Sonnet, LLaMA, and DeepSeek could accept spreadsheet inputs, they were limited to textual outputs and failed to reliably produce tabular data with the expected number of rows, even when explicitly prompted to do so.

Additional model-specific constraints further reduced feasibility: Mistral did not support document inputs on the tested endpoint, while models in the LLaMA family repeatedly exceeded the document ingestion time limit. Consequently, we abandoned the Bedrock-based setup and re-ran the experiments using each model’s publicly available chatbot interface. Grok was unavailable in Italy at the time of the experiment and was therefore excluded. Among the remaining models, Kimi~K2 Thinking failed to consistently generate valid tabular outputs, producing either an incorrect number of entries or malformed files.

For GPT, DeepSeek, and Qwen, we successfully prompted the generation of spreadsheet files. This approach did not generalize to Claude, which instead produced a non-functional interactive web application. However, requesting tabular data as structured lists yielded outputs with the correct number of entries, allowing their inclusion in subsequent analyses.

\paragraph{Case studies}\label{sec:casestudies}
To investigate the sources of classification errors, we qualitatively analyzed a subset of LLM-generated annotations. This analysis revealed recurring patterns underlying the models’ limited performance. For example, in prompt~3, ChatGPT-5 systematically labeled the \textit{Variable/Method identifiers are mentioned} feature as present for all annotators, regardless of comment content.

A common challenge across models was distinguishing general technical terminology from code-specific identifiers. Some models appeared overly sensitive to keywords (e.g., GPT labeling all occurrences of “vulnerable” as positive), while others were misled by generic technical terms such as “debug” (e.g., Qwen). In general, it seems models can get confused by the literal meaning of words and miss the high-level significance, for example "By elimination it seems that there is no other lines that can be assessed as vulnerables" for human annotators was just positive for \textit{I don't find vulnerability}, but DeepSeek also marked it as positive for \textit{Text is related to the specific vulnerability}. Models also varied in their interpretation of uncertainty; for instance, GPT and Claude occasionally inferred uncertainty from modal expressions that human annotators did not consider indicative of uncertainty ("I would be inclined to think" or "it is likely that"). Finally, certain misclassifications remained difficult to explain, for example, DeepSeek marked "I am not very sure it is a vulnerable line" as positive for \textit{Relevant Keywords mentioned}, but "I don't think any of the lines are vulnerable" as negative, even if they include almost all the same words.

\section{Discussion} \label{sec:discussion:threats}

%\subsection{Key Negative Findings and Insights}
%\yg{please check again after getting the final results}

Our exploration revealed both the potential and the limits of using LLMs for security-specific annotations of subjects' justifications for security choices in experiments where humans are asked to identify security issues in code snippets. 

While the model demonstrated the ability to parse structured input and apply simple binary classification tasks, its performance varied widely across codes, especially those involving nuanced reasoning or implicit context. Most importantly, it is not possible to achieve even a moderate accuracy of the given human annotators. Therefore, the use of LLMs as an automatic annotator is still not recommended for security-relevant annotations.

Prompt engineering showed diminishing returns. Adding detailed explanations and examples improved results marginally, but not to a degree that would significantly offset the manual effort required to craft such prompts.
% Interestingly, attempts to guide the model interactively—by identifying its mistakes, providing focused clarifications, or answering its own questions—did not lead to sustained accuracy gains. In some cases, performance worsened.
Despite increased prompt complexity and refinement, performance improvements were limited. 
% In particular, iterative steps such as highlighting the most error-prone codes or prompting the model to emulate human annotation decisions failed to yield substantial gains. Even the most interactive approach—asking the model to identify “controversial” examples and submit clarification questions—resulted in lower overall accuracy.
These outcomes indicate that the model’s initial limitations in understanding security-specific annotation tasks were not easily overcome by progressively more elaborate prompting alone.

These findings suggest the following key insights:
\begin{itemize}
    \item (some) LLMs can assist with straightforward code detection (e.g., mentions of line numbers or identifiers), but struggle with interpretive codes like confusion.
    \item Prompt quality matters, but beyond a point, iterative refinement without architectural adaptation (e.g., fine-tuning) may yield diminishing returns.
    \item Interactive prompting strategies may not be sufficient to bridge the semantic gap between surface-level understanding and security-informed reasoning. 
\end{itemize}

\section{Related Work} \label{sec:related:work}

\noindent\textbf{Opinion mining for Software development}
Recent research has increasingly applied opinion mining and sentiment analysis to understand developers' perspectives across various software engineering (SE) tasks \cite{lin2022opinion}. 
Various Natural Language Processing (NLP) and Machine Learning (ML) techniques have been applied to analyze qualitative text in software engineering contexts. 
Sentiment polarity detection—using lexicon-based tools like SentiStrength \cite{thelwall2010sentiment}, ML models like Senti4SD \cite{calefato2018sentiment}, and Deep Learning classifiers like SentiCR \cite{ahmed2017senticr}—has been used to classify developer emotions in issue reports, Stack Overflow posts, and code reviews. 
Topic modeling methods such as LDA \cite{blei2003latent} and TwitterLDA \cite{zhao2011comparing} have supported the discovery of latent concerns in developer discussions and user reviews. 
Supervised classifiers have been employed to extract structured insights from unstructured text, such as identifying bug reports, code requests, or usability issues in app reviews (e.g., MARC 3.0 \cite{jha2018using, jha2019mining}, Ticket-Tagger \cite{kallis2019ticket}). 
More recently, hybrid techniques combining rule-based and statistical models have been used for emotion detection (e.g., DEVA \cite{islam2018deva}, EmoTxT \cite{calefato2017emotxt}), and trust inference in developer collaboration \cite{da2016estimating}. 
Despite these advances, many studies report performance limitations when applying tools trained on general-domain data to software-specific texts, underscoring the need for domain-adapted approaches and annotated datasets curated for SE tasks \cite{jongeling2017negative}.

\noindent\textbf{Qualitative data analysis in SE experiments.}
Qualitative data analysis plays a vital role in SE experiments by uncovering insights from textual sources such as interviews, open-ended surveys, and observational data \cite{wohlin2012experimentation}. 
The goal is to generate credible findings while maintaining a transparent chain of evidence, linking conclusions clearly to the original data \cite{runeson2009guidelines}. 
The process involves coding segments of text to capture recurring ideas and developing themes or hypotheses through techniques like constant comparison or cross-case analysis \cite{seaman1999qualitative}. 
These are followed by hypothesis confirmation strategies, such as triangulation or replication.
To reduce bias, multiple researchers often code independently and reconcile results collaboratively. 
These methods have supported research into developer rationale in code review comments \cite{mantyla2008types}, selection of patches for vulnerability repair \cite{papotti2024acceptance},  vulnerability identification with program slicing \cite{papotti2025effects}, barriers to adopting automated tools \cite{johnson2016cross}, and developer sentiments in socio-technical systems \cite{ford2016paradise}. Such studies highlight the importance of rigorous qualitative analysis for exploring the human and collaborative aspects of software engineering.

\noindent\textbf{LLMs for qualitative data analysis.}
% \todo[inline]{MC: Expectations from similar setups to compare to what we found here}
With the development of large language models, their potential to assist humans in qualitative research has been discussed increasingly. Several studies have explored the application of LLMs in qualitative analysis across different domains. For example, in healthcare, GPT-4 demonstrates moderate agreement with human qualitative analysis~\cite{li2024comparing}; in psychology, LLMs have been tested for deductive coding tasks in analyzing children’s curiosity-driven questions, showing moderate to high agreement with experts on certain metrics~\cite{xiao2023supporting}; in processing, LLMs have also been applied for historical literature annotation~\cite{dunivin2024scalable}, music shuffle preference annotation and password management annotation~\cite{dai2023llm}, and show the potential ability. In software engineering, the usage of LLMs for qualitative data analysis is a relatively new topic. \cite{rasheed2024can} proposes a multi-agent strategy to study the LLMs' ability in different qualitative research tasks, involving thematic annotation for GitHub or Stack Overflow discussions. While showing potential in qualitative research, some limitations of LLMs have also raised concerns, such as model biases, illusions, and ethical problems~\cite{schroeder2025large}. However, LLMs' ability to analyze technical comments and to annotate different types of themes remains to be explored.

\section{Threats to Validity}

% \todo[inline]{MC: The impact of prompt engineering (goes to the third paragraph)}

\paragraph{Selected task might not represent all annotation scenarios}
Our evaluation focuses on a single type of security experiment and a fixed set of nine security-relevant codes. The comments dataset is relatively representative of vulnerability assessment comments, coming from CS master's students. However, the tasks and code can be different in the scope of security-specific annotation. For instance, the comments might cause bias since the students' comments can vary a lot from those of the developers in the real-world. Moreover, in this paper, the task is for LLMs strictly as full replacements for human annotators, while the ability of LLMs to suggest candidate code, pre-filter, or detect disagreement remains to be defined.

\paragraph{Selected LLMs might not represent future performance}
The selection of large language models was guided by the widely used benchmark LiveBench. All LLMs were utilized in their full versions under subscription plans. However, the models were chosen at a specific point in time and may not reflect future developments. 

\paragraph{The impact of prompt engineering}
Our prompt design aimed to steer LLM outputs toward reliable annotations by embedding clear task descriptions, structured guidelines, and representative examples, thus leveraging in‑context learning and few‑shot prompting to reduce ambiguity. Additionally, we iteratively refined prompts,\promptthree in particular, through pilot interactions with GPT‑4o, adjusting phrasing and examples based on model behavior to mitigate obvious misinterpretations. These practices align with common prompt engineering techniques such as template structuring and exemplar selection identified in recent literature on prompting methodologies \cite{marvin2023prompt} and systematic taxonomies of prompting strategies \cite{schulhoff2024prompt}. Nonetheless, prompt engineering lacks well‑established, universally optimal procedures, and the effectiveness of particular formulations can vary substantially across tasks and models. Our design choices, therefore, reflect subjective decisions and limited expertise, and alternative prompt structures could lead to different model behaviors, limiting generalizability and posing a threat to validity.

% \paragraph{Internal Validity}
% During the data collection stage, annotators worked independently on the initial comments to produce individual annotations, following the procedures of the annotation task. When the annotation procedure requires separate work, annotators cannot influence each other’s justifications. For the LLMs under evaluation, identical prompts were provided to ensure consistency. 
% \paragraph{External Validity}
% \paragraph{Construct validity}
% Since our experiments are conducted on unbalanced datasets, to avoid some performance metrics being influenced by chance, we use change-correct accuracy as the primary metric in our study. This metric is adjusted by subtracting the expected values that would be obtained by chance.
% \paragraph{Conclusion Validity}
% The results in this paper are supported by statistical tests.To compare the effectiveness of prompts or LLMs, the samples (change corrected accuracy Acc$^*$) are paired—for instance, measuring the improvement in accuracy of the same code under two different prompts. Therefore, applying the Wilcoxon paired test is appropriate to address our research questions. The sample sizes of both datasets are large enough to support reliable conclusions. The Bonferroni correction is applied to adjust the significance level in multiple comparisons, reducing the risk of false positive conclusions.

\section{Conclusion} \label{sec:conclusions}

Recent research has shown that large language models (LLMs) can replicate human annotators for extracting sentiment analysis from text. In this paper, we have investigated whether code capturing domain-specific aspects in security and software engineering, such as \texttt{code identifiers mentioned}, \texttt{lines-of-code-mentioned}, \texttt{security keywords mentioned}, can be achieved by LLMs to reduce the manual effort involved in the qualitative analysis of technical comments by acting as automated annotators.

We have prompted the four best proprietary and open source LLMs on LiveBench (\gptmodel,  \claudemodel, \modelthree, and \modelfour) to identify the presence of 9 security-relevant codes in free-text comments from humans analyzing code snippets for vulnerabilities. The LLMs' outputs were compared against a ground truth annotated by expert coders using precision, recall, and the Heidke Skill Score (a chance-corrected accuracy measure). We refined the prompts by mimicking the process of human annotators: emerging codes, a codebook with examples, and conflicting examples.

We observed small improvements after providing detailed descriptions and examples for each code, but we did not see a statistically significant gain across codes or models. Most importantly, the overall performance is too low to replace a human annotator.
% as its chance corrected accuracy is lower than 40\%.

Given these findings, we argue that the role of LLMs in qualitative security coding should, at present, remain assistive rather than autonomous. More experiments on larger corpora of experiments are needed to go beyond generating a first draft of annotations, highlighting potential codes for human verification, or accelerating low-complexity tagging tasks.

These results motivate future work on structured prompting interfaces, model calibration, and the design of hybrid pipelines where LLMs complement rather than replace human insight in technically demanding qualitative analyses.

\section*{Acknowledgments}
This work was partly funded by the EU under the Horizon Europe Program with n.  101120393 (\href{https://sec4ai4sec.eu}{Sec4AI4Sec}), by the Italian Ministry of University and Research (MUR) under the P.N.R.R. – NextGenerationEU grant n.\ PE00000014 (SERICS subproject, CUP E63C24000590001).

\section*{CRediT}

%\begin{description}
\emph{Conceptualization:} MC, YG, FM;
% Ideas; formulation or evolution of overarching research goals and aims
\emph{Methodology:} MC, YG, FM;
%Development or design of methodology; creation of models
\emph{Software Programming:} MC;
%software development; designing computer programs; implementation of the computer code and supporting algorithms; testing of existing code components
\emph{Validation:} YG, FM;
%Verification, whether as a part of the activity or separate, of the overall replication/ reproducibility of results/experiments and other research outputs
\emph{Formal analysis:} FM;
% Application of statistical, mathematical, computational, or other formal techniques to analyze or synthesize study data
\emph{Investigation:} MC, YG;
%Conducting a research and investigation process, specifically performing the experiments, or data/evidence collection
\emph{Resources:} FM;
%Provision of study materials, reagents, materials, patients, laboratory samples, animals, instrumentation, computing resources, or other analysis tools
\emph{Data Curation:} MC;
%Management activities to annotate (produce metadata), scrub data and maintain research data (including software code, where it is necessary for interpreting the data itself) for initial use and later reuse
\emph{Writing - Original Draft:} MC, YG;
%Preparation, creation and/or presentation of the published work, specifically writing the initial draft (including substantive translation)
\emph{Writing - Review \& Editing:} %MC, YG, FM;
%Preparation, creation and/or presentation of the published work by those from the original research group, specifically critical review, commentary or revision – including pre-or postpublication stages
\emph{Visualization:} MC , YG;
% Preparation, creation and/or presentation of the published work, specifically visualization/ data presentation
\emph{Supervision:} YG, FM;
%Oversight and leadership responsibility for the research activity planning and execution, including mentorship external to the core team
\emph{Project administration:} FM;
%Management and coordination responsibility for the research activity planning and execution
\emph{Funding acquisition:} FM.
%Acquisition of the financial support for the project leading to this publication 

\clearpage